\documentclass[twocolumn]{aastex63}
\usepackage{booktabs}
\usepackage{amsmath}

 \newcommand{\E}[1]{\ensuremath{\times 10^{#1}}}

\def\gtrsim{\mathrel{\hbox{\rlap{\hbox{\lower4pt\hbox{$\sim$}}}\hbox{$>$}}}}

\def \arcsec {\hbox{$^{\prime\prime}$}}

\received{\today}
\revised{--}
\accepted{--}

\submitjournal{ApJ}

\shorttitle{1RXS J165424.6-433758}
\shortauthors{O'Connor et al.}
\graphicspath{{./}{figures/}}

\begin{document}

\title{Identification of 1RXS J165424.6-433758 as a polar cataclysmic variable}

\correspondingauthor{Brendan O'Connor}
\email{oconnorb@gwmail.gwu.edu}

\author[0000-0002-9700-0036]{B. O'Connor}
   \affiliation{Department of Physics, The George Washington University, Washington, DC 20052, USA}
    \affiliation{Astronomy, Physics and Statistics Institute of Sciences (APSIS), The George Washington University, Washington, DC 20052, USA}
    \affiliation{Department of Astronomy, University of Maryland, College Park, MD 20742-4111, USA}
    \affiliation{Astrophysics Science Division, NASA Goddard Space Flight Center, 8800 Greenbelt Rd, Greenbelt, MD 20771, USA}
\author[0000-0003-0030-7566]{J. Brink}
    \affiliation{Department of Astronomy, University of Cape Town, Private Bag X3, Rondebosch 7701, South Africa}
    \affiliation{South African Astronomical Observatory, P.O. Box 9, Observatory 7935, Cape Town, South Africa}
\author[0000-0002-7004-9956]{D.~A.~H. Buckley}
    \affiliation{South African Astronomical Observatory, P.O. Box 9, Observatory 7935, Cape Town, South Africa}
    \affiliation{Southern African Large Telescope, P.O. Box 9, Observatory 7935, Cape Town, South Africa}
    \affiliation{Department of Astronomy, University of Cape Town, Private Bag X3, Rondebosch 7701, South Africa}
    \affiliation{Department of Physics, University of the Free State, P.O. Box 339, Bloemfonein 9300, South Africa}
\author[0000-0002-8286-8094]{K. Mukai} 
    \affiliation{CRESST II and X-ray Astrophysics Laboratory, NASA/GSFC, Greenbelt, MD 20771, USA}
    \affiliation{Department of Physics, University of Maryland Baltimore County, 1000 Hilltop Circle, Baltimore MD 21250, USA}
\author[0000-0003-1443-593X]{C. Kouveliotou}
    \affiliation{Department of Physics, The George Washington University, Washington, DC 20052, USA}
    \affiliation{Astronomy, Physics and Statistics Institute of Sciences (APSIS), The George Washington University, Washington, DC 20052, USA}
\author[0000-0002-5274-6790]{E. G\"{o}\u{g}\"{u}\c{s}}
    \affiliation{Sabanc\i~University, Faculty of Engineering and Natural Sciences, \.Istanbul 34956, Turkey}
\author[0000-0002-5956-2249]{S.~B. Potter}
    \affiliation{South African Astronomical Observatory, P.O. Box 9, 7935 Observatory, South Africa}
    \affiliation{Department of Physics, University of Johannesburg, PO Box 524, Auckland Park 2006, South Africa}
\author[0000-0002-6896-1655]{P. Woudt}
    \affiliation{Department of Astronomy, University of Cape Town, Private Bag X3, Rondebosch 7701, South Africa}
\author[0000-0002-7851-9756]{A. Lien} 
    \affiliation{University of Tampa, Department of Chemistry, Biochemistry, and Physics, 401 W. Kennedy Blvd, Tampa, FL 33606, USA}
\author[0000-0001-7821-9369]{A. Levan}
    \affiliation{Department of Astrophysics/IMAPP, Radboud University, P.O. Box 9010, 6500 GL, The Netherlands}
\author[0000-0002-6447-4251]{O. Kargaltsev}
    \affiliation{Department of Physics, The George Washington University, Washington, DC 20052, USA}
   \affiliation{Astronomy, Physics and Statistics Institute of Sciences (APSIS), The George Washington University, Washington, DC 20052, USA}
\author[0000-0003-4433-1365]{M.~G. Baring}
    \affiliation{Department of Physics and Astronomy - MS 108, Rice University, 6100 Main Street, Houston, Texas 77251-1892, USA}
\author[0000-0001-8018-5348]{E. Bellm} 
    \affiliation{DIRAC Institute, Department of Astronomy, University of Washington, 3910 15th Avenue NE, Seattle, WA 98195, USA}
\author[0000-0003-1673-970X]{S.~B. Cenko}
    \affiliation{Astrophysics Science Division, NASA Goddard Space Flight Center, 8800 reenbelt Rd, Greenbelt, MD 20771, USA}
    \affiliation{Joint Space-Science Institute, University of Maryland, College Park, MD 20742 USA}
\author[0000-0002-8465-3353]{P.~A. Evans}
    \affiliation{School of Physics and Astronomy, University of Leicester, University Road, Leicester, LE1 7RH, UK}
\author[0000-0001-8530-8941]{J. Granot}
    \affiliation{Department of Natural Sciences, The Open University of Israel, P.O Box 808, Ra'anana 43537, Israel}
    \affiliation{Astrophysics Research Center of the Open university (ARCO), The Open University of Israel, P.O Box 808, Ra’anana 43537, Israel}
    \affiliation{Department of Physics, The George Washington University, Washington, DC 20052, USA}
\author{C. Hailey}
    \affiliation{Columbia Astrophysics Laboratory, Columbia University, New York, NY 10027, USA}
\author[0000-0003-2992-8024]{F. Harrison}
    \affiliation{Cahill Center for Astrophysics, California Institute of Technology, 1216 East California Boulevard, Pasadena, CA 91125, USA}
\author[0000-0002-8028-0991]{D. Hartmann}
    \affiliation{Department of Physics and Astronomy, Clemson University, Kinard Lab of Physics, Clemson, SC 29634-0978, USA} 
\author[0000-0001-9149-6707]{A.~J. van der Horst}
    \affiliation{Department of Physics, The George Washington University, Washington, DC 20052, USA}
     \affiliation{Astronomy, Physics and Statistics Institute of Sciences (APSIS), The George Washington University, Washington, DC 20052, USA}
\author[0000-0002-1169-7486]{D. Huppenkothen}
    \affiliation{SRON Netherlands Institute for Space Research, Niels Bohrweg 4, 2333CA Leiden, The Netherlands}
\author{L. Kaper}
    \affiliation{University of Amsterdam, Science Park 904, 1098 XH Amsterdam, The Netherlands}
\author[0000-0002-6745-4790]{J.~A. Kennea}
    \affiliation{Department of Astronomy and Astrophysics, The Pennsylvania State University, 525 Davey Lab, University Park, PA 16802, USA}
\author[0000-0002-6986-6756]{P.~O. Slane}
    \affiliation{Center for Astrophysics, Harvard \& Smithsonian, 60 Garden St. Cambridge, MA 02138, USA}
\author[0000-0003-2686-9241]{D. Stern}
    \affiliation{Jet Propulsion Laboratory, California Institute of Technology, 4800 Oak Grove Drive, Mail Stop 169-221, Pasadena, CA 91109, USA}
\author[0000-0002-1869-7817]{E. Troja}
    \affiliation{University of Rome Tor Vergata, Department of Physics, via della Ricerca Scientifica 1, 00100, Rome, IT}
\author[0000-0002-3101-1808]{R.~A.~M.~J. Wijers}
    \affiliation{Anton Pannekoek Institute, University of Amsterdam, Postbus 94249, 1090 GE Amsterdam, The Netherlands}
    \affiliation{Department of Physics, The George Washington University, Washington, DC 20052, USA}
\author[0000-0002-7991-028X]{G. Younes}
    \affiliation{Department of Physics, The George Washington University, Washington, DC 20052, USA}
    \affiliation{Astronomy, Physics and Statistics Institute of Sciences (APSIS), The George Washington University, Washington, DC 20052, USA}



\begin{abstract}
We present the results of our X-ray, ultraviolet, and optical follow-up campaigns of 1RXS J165424.6-433758, an X-ray source detected with the \textit{Swift} Deep Galactic Plane Survey (DGPS). The source X-ray spectrum (\textit{Swift} and \textit{NuSTAR}) is described by thermal bremsstrahlung radiation with a temperature of $kT=10.1\pm1.2$ keV, yielding an X-ray ($0.3-10$ keV) luminosity $L_X=(6.5\pm0.8)\times10^{31}$ erg s$^{-1}$ at a \textit{Gaia} distance of 460 pc. Spectroscopy with the Southern African Large Telescope (SALT) revealed a flat continuum dominated by emission features, demonstrating an inverse Balmer decrement, the $\lambda4640$ Bowen blend, almost a dozen HeI lines, and HeII $\lambda4541$, $\lambda4686$ and $\lambda 5411$. Our high-speed photometry demonstrates a preponderance of flickering and flaring episodes, and revealed the orbital period of the system,  $P_\textrm{orb}=2.87$ hr, which fell well within the cataclysmic variable (CV) period gap between $2-3$ hr. These features classify 1RXS J165424.6-433758 as a nearby polar magnetic CV.
\end{abstract}



\section{Introduction}
\label{sec: intro}


Cataclysmic Variables (CVs) are binary systems composed of a white dwarf (WD) and a late-type main sequence star. An accretion disk will also form if the WD magnetic field is $<$\,$10^6$ G \citep{Warner1995}. In systems with a highly magnetic WD ($\sim$\,$10^{6-9}$ G) the accretion disk is either fully (\citealt{Cropper1990}, e.g., AM Herculis; a polar), or partially disrupted (\citealt{Patterson1994}, e.g., DQ Herculis; an intermediate polar (IP)).

In polar magnetic CVs (mCVs), matter from the companion star accretes onto the WD poles following the magnetic field lines. As the matter in the accretion column collides with the WD atmosphere a shock is formed, which emits cyclotron (optical/infrared) and thermal bremsstrahlung (X-rays) radiation. Polars have typical soft X-ray ($0.3-10$ keV) luminosities of $\sim$\,$10^{30-32}$ erg s$^{-1}$. 

In these systems, the large magnetic field ($>$\,$10^7$ G) causes synchronicity with the companion star, such that the orbital period and WD rotational period are the same. Angular momentum losses lead to an evolution of the orbital period over time. In CVs with large orbital periods ($>$\,$3$ hr), the orbit is dominated by magnetic breaking, while in short period binaries ($<$\,$2$ hr) gravitational radiation losses are the dominant factor. However, magnetic breaking is thought to be ineffective when the companion star becomes fully convective, leading to a so called ``period gap'' between $2$\,$-$\,$3$ hr \citep{Knigge2006}. Despite this, many polars  are discovered with orbital periods lying in this gap, requiring new theories of mCV formation and evolution (e.g., magnetic field evolution or reduced magnetic braking; \citealt{Li1994,Belloni2020,Webbink2002}).


The disruption of the accretion disk reduces the thermal instabilities in these systems, making novae and outbursts less common than in non-magnetic CVs. As a result, magnetic CVs are rare in optical surveys \citep{Oliveira2017,Oliveira2020}. For example, \citet{Szkody2020} found that only $\sim$\,$1/100$ CVs discovered in optical surveys are magnetic. However, using a volume limited ($<$\,$150$ pc) sample, \citet{Pala2020} showed that 36\% of CVs, selected using \textit{Gaia} parallaxes, are magnetic. This emphasizes the need for X-ray surveys (e.g., \textit{e-ROSITA}) to improve the discovery rate of mCVs \citep[see also][]{Bernardini2012}.


In this work, we introduce the discovery of a new polar through the \textit{Swift} Deep Galactic Plane Survey (DGPS; \citealt{DGPS}), a \textit{Swift} Key Project and \textit{NuSTAR} Legacy Project. The DGPS covers 40 deg$^{2}$ of the Galactic Plane between $10$\,$<$\,$|l|$\,$<$\,$30$ deg in longitude and $|b|$\,$<$\,$0.5$ deg in latitude, to luminosity $L_X$\,$>$\,$10^{34}$ erg s$^{-1}$, out to 3-6 kpc. The DGPS has also revealed the discovery of a distant IP CV \citep{Gorgone2021} and a candidate IP \citep{J1708}, totaling three mCVs serendipitously discovered by the survey.

1RXS J165424.6-43375 (hereafter J1654) was discovered with \textit{ROSAT} in 1990 August, and later with \textit{ASCA} as AX J165420-4337. The Chasing the Identification of ASCA Galactic Objects (ChIcAGO) survey identified the X-ray counterpart to AX J165420-4337 using \textit{Chandra} \citep{Anderson2014}. The source was later observed with \textit{Swift} \citep{Reynolds2013} as part of a shallow Galactic plane survey.

\cite{Takata2022} analyzed a sample of unclassified X-ray sources with \textit{Gaia} counterparts, including J1654 (G596 in that paper). Using a combination of \textit{Swift}, \textit{NuSTAR}, and the Transiting Exoplanet Survey Satellite (TESS), they classify J1654 as a candidate polar. In particular, their result is due to the periodicity of 2.88 hr uncovered by TESS, and the fact that no other periodicity is identified, suggesting the spin and orbital periods may be locked. However, they do not exclude the possibility that J1654 is a IP.

Here, we present DGPS observations and follow-up of J1654 obtained using our approved Target of Opportunity (ToO) programs. We obtained observations with \textit{Swift}, \textit{NuSTAR}, the Southern African Large Telescope (SALT), and the South African Astronomical Observatory (SAAO) 1-m telescope. We also analyze archival optical and infrared observations, and archival \textit{XMM-Newton} observations. We uncover clear low- and high-state transitions in the ultraviolet, optical, infrared, and X-rays. The low-state photometry allows us to analyze the secondary star, determining that it is likely of late spectral type. High-speed photometry carried out over multiple nights yields an orbital period of $\sim$2.87 hr, in agreement with \cite{Takata2022}. Our optical spectroscopy reveals numerous lines of hydrogen and helium, demonstrating an inverse Balmer decrement. In particular, our spectroscopic observations point towards the classification of the source as a typical polar at a \textit{Gaia} distance of $\sim$460 pc, solidifying its previous classification \citep{Takata2022}.

We present the observations and analysis in \S \ref{sec: obs/analysisJ1654}. The results of our timing and spectroscopic analyses are presented in \S \ref{sec: resultsJ1654} with a brief discussion in \S \ref{sec: discussionJ1654}, and conclusions in \S \ref{sec: conclusionsJ1654}. Throughout the manuscript errors are quoted at the $1\sigma$ level and upper limits at the $3\sigma$ level, unless otherwise stated.

\section{Observations and Data Analysis}
\label{sec: obs/analysisJ1654}

We begin by presenting our ultraviolet, optical and infrared data analysis in \S \ref{sec: optIRanalysisJ1654}, including optical spectroscopy (\S \ref{sec: saltJ1654}) and high-speed photometry (\S \ref{sec: saaoJ1654}). This is followed by the description of our X-ray data analysis in \S \ref{sec: XrayanalysisJ1654}.

\subsection{Ultraviolet/optical/infrared Data}
\label{sec: optIRanalysisJ1654}

\begin{figure*}
\centering
\includegraphics[width=1.8\columnwidth]{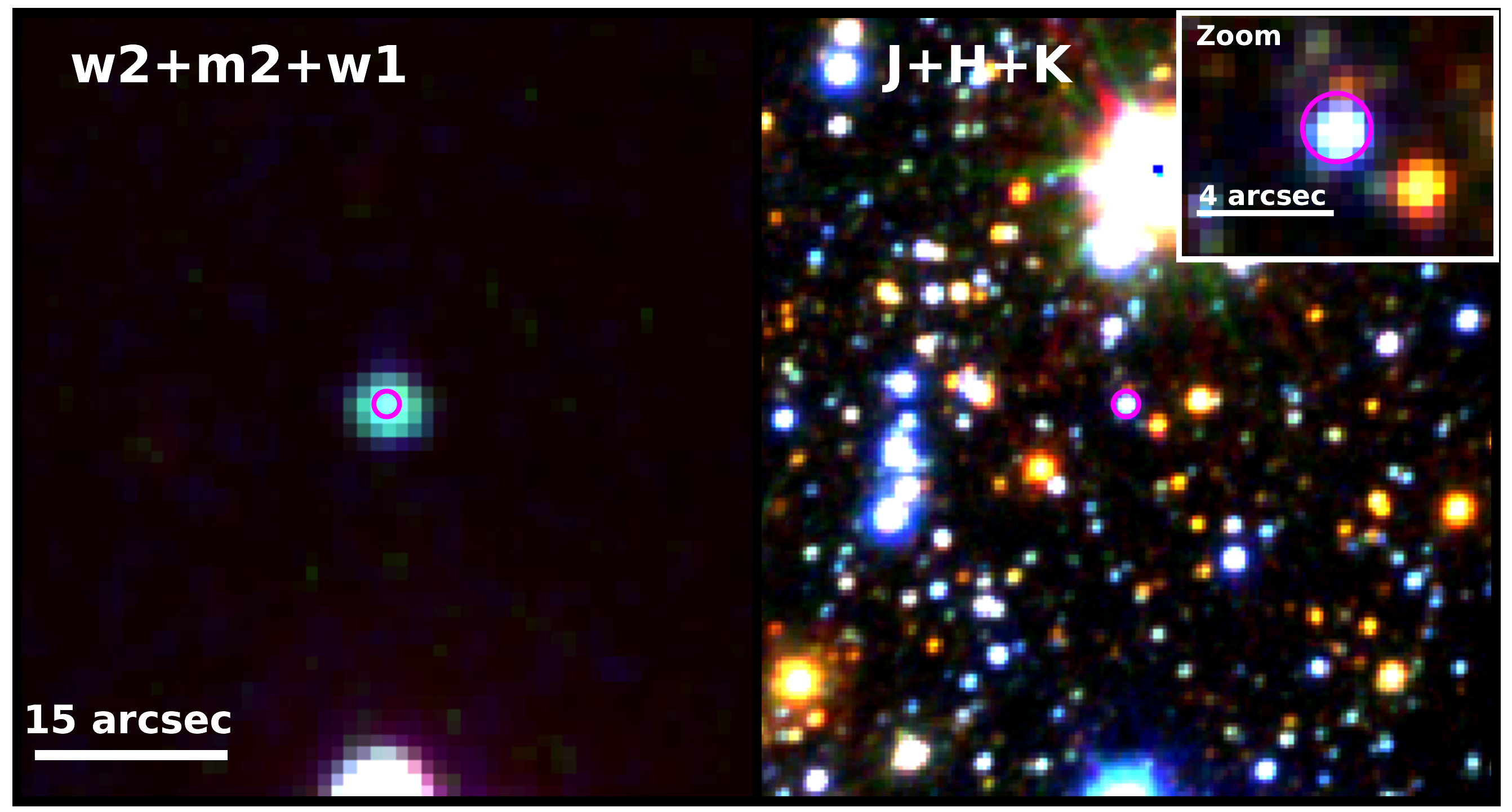}
\caption{RGB image of the field of J1654 using three UVOT filters (\textit{\textbf{left}}: red = $uvw1$, green = $uvm2$, and blue = $uvw2$) and three VVV filters (\textit{\textbf{right}}: red = $K_s$, green = $H$, and blue = $J$). The archival \textit{Chandra} localization ($1.0\arcsec$; 95\% CL) is displayed by the magenta circle. In the top right a zoom in on the position of the counterpart is shown. North is up and East is to the left. 
}
\label{fig: counterpartJ1654}
\end{figure*}

\subsubsection{Archival Multi-wavelength Data}

Throughout the paper, we utilize the archival \textit{Chandra} localization of J1654 at RA, DEC (J2000) = $16^{h}54^m 23^{s}.45$, $-43^\circ 37\arcmin 43.8\arcsec$ with uncertainty of $0.43\arcsec$ \citep[68\% CL;][]{Evans2010,Anderson2014}.  
This position is covered by multiple archival surveys, including \textit{Gaia} \citep{Gaia2016,Gaia2018,gaiaEDR3}, the VISTA Variables in the V\'{i}a L\'{a}ctea (VVV) Infrared Astrometric Catalogue \citep[VIRAC;][]{Smith2018}, the VST 
Photometric H$\alpha$ Survey of the Southern Galactic Plane and Bulge \citep[VPHAS;][]{Drew2014,Drew2016}, the Asteroid Terrestrial-impact Last Alert System (ATLAS) reference catalog \citep{Tonry2021}, and the Galactic Legacy Infrared Mid-Plane Survey Extraordinaire \citep[GLIMPSE;][]{Benjamin2003}. In these catalogs we identify a bright optical and infrared counterpart coincident with the \textit{Chandra} localization (Figure \ref{fig: counterpartJ1654}). We tabulate archival photometry from these catalogs in Table \ref{tab: optcounterpartJ1654}.

The \textit{Gaia} counterpart yields a nearby distance $d=463\pm24$ pc 
\citep{Bailer-Jones2021} for J1654. 
Moreover, the counterpart has a high proper motion of $\mu_\textrm{RA}=22.2\pm0.3$ mas yr$^{-1}$ and $\mu_\textrm{DEC}=-7.3\pm0.3$ mas yr$^{-1}$. However, this proper motion is not high enough to have impacted the source's X-ray localization over the past 20 years by more than $\sim$0.5\arcsec.

We also performed an analysis of public archival data obtained in multiple epochs. 
We retrieved calibrated VPHAS images from the ESO Archive Science Portal\footnote{\url{https://archive.eso.org/scienceportal/home}}. VPHAS photometry was performed in two epochs: 2013 July 27 in $ri$(H$\alpha$) and 2015 April 23 in the $ugr$ filters. 
We performed aperture photometry using \texttt{SExtractor} with an aperture radius of $1.5\times$ the image full width at half maximum (FWHM). The photometry was calibrated to VPHAS DR2 for the $ri$ filters and ATLAS for the $g$ filter. The photometry derived on 2013 July 27 agrees with that presented in VPHAS DR2, but on 2015 April 23 the source is brighter by $\sim$2 magnitudes, which is indicative of a high-state vs low-state transition \citep{Warner1999,Hessman2000,Wu2008}. The photometry is reported in Table \ref{tab: optcounterpartJ1654}.

We performed a similar analysis of the archival VVV data to search for state changes. The calibrated images were retrieved from the ESO Archive Science Portal. Observations covering the field of J1654 were obtained in the $ZYJHK$ filters. 
The $Z$-band photometry covers five epochs, whereas $YJH$ images were obtained only in two epochs on 2014 November 12 and 2017 January 18. 
We performed photometry using \texttt{SExtractor}, and determined the photometric zeropoint by calibrating to point sources in the VIRAC catalog \citep{Smith2020} for the $ZYJH$ bands and VVV DR2 \citep{Minniti2017} for the $K$-band. Based on the $Z$-band variability ($\Delta m_\lambda\sim1.1$ mag), we find that the observations obtained on 2017 January 18 are in a high-state, and those from 2014 November 12 are in a low-state (see Table \ref{tab: optcounterpartJ1654}).

We further searched images obtained by the \textit{Spitzer Space Telescope} (hereafter \textit{Spitzer}) through GLIMPSE. PSF photometry was performed with \texttt{SExtractor}, calibrated to GLIMPSE standards in the field. The results are shown in Table \ref{tab: optcounterpartJ1654}.

\subsubsection{Ultraviolet Observations}

\label{sec: uvotJ1654}

Data obtained with the \textit{Swift} UltraViolet and Optical Telescope \citep[UVOT;][]{Roming2005,Breeveld2011} were analyzed using \texttt{HEASoft v6.29c} with \texttt{CALDB} version 20211108. The counterpart was detected in all single UVOT exposures in the $uvw2$, $uvm2$, $uvw1$, and $u$ filters. A finding chart is shown in Figure \ref{fig: counterpartJ1654}.  
We performed aperture photometry using the \texttt{uvotsource} task with a circular source aperture of radius  $5\arcsec$ and a nearby circular background region of radius $15\arcsec$. The photometry is reported in Table \ref{tab: observationsUVOT} in the AB magnitude system.

We likewise analyzed archival data obtained with the \textit{XMM-Newton} Optical/UV  Monitor (OM) Telescope \citep{Mason2001}. Observations were carried out in the $UVM2$, $UVW1$, $B$, and $V$ filters. The source is detected ($\sim5\sigma$) only in the $UVW1$ filter, which has the longest exposure time (5000 s). We used the \texttt{omichain} command to reduce the data. We used \texttt{omdetect} to determine the source count rates and the \texttt{ommag} task to determine the instrumental magnitudes. We converted the instrumental magnitudes to the AB system using the standard zeropoints\footnote{\url{https://xmm-tools.cosmos.esa.int/external/xmm\_user\_support/documentation/uhb/omfilters.html}}.

\subsubsection{SALT Optical Spectroscopy}
\label{sec: saltJ1654}

Longslit spectroscopic observations were conducted using the 10-m class Southern African Large Telescope \citep[SALT;][]{Buckley2006} equipped with the Robert Stobie Spectrograph \citep[RSS;][]{Burgh2003,Kobulnicky2003,Smith2006}. An initial spectrum was obtained on 2022 April 26 using the low resolution PG0300 grating at a grating tilt of $5.75^\circ$ with slit width $1.25\arcsec$. The spectrum covered the wavelength range 3200 \AA\ to 9000 \AA\ with a resolution of $R\approx400$ at the central wavelength (6210 \AA). Additional frame-transfer spectroscopic observations were conducted on 2022 May 2, 10, and 27 using the high resolution PG0900 grating at a grating tilt $14.75^\circ$ with a slit width of $1.25\arcsec$. The frame-transfer spectra covered 4060 \AA\ to 7120 \AA\ with a resolution of $R\approx1200$ at the central wavelength (5620 \AA). The exposure time for the frame-transfer spectra is 100 s. A log of SALT observations is provided in Table \ref{tab: observationsSALT}.

The spectra were reduced using the \texttt{PySALT} package \citep{Crawford2010}, which includes bias subtraction, flat-fielding, amplifier mosaicing, and a process to remove cosmetic defects. The spectra were wavelength calibrated, background subtracted, and extracted using standard IRAF\footnote{IRAF is distributed by the National Optical Astronomy Observatory, which is operated by the Association of Universities for Research in Astronomy (AURA) under cooperative agreement with the National Science Foundation (NSF).} procedures. We obtained a relative flux calibration of all spectra using spectrophotometric standard stars. For the observation obtained on 2022 April 26, we used the standard star HILT600, and for the frame-transfer observations we used the star LTT4364. The frame-transfer observations were further continuum normalized, and used to create trailed spectra. The spectra are discussed in \S \ref{sec: OptspecanalysisJ1654}.

\subsubsection{SAAO High-speed Photometry}
\label{sec: saaoJ1654}

We obtained high-speed photometry using the Sutherland High speed Optical Camera \citep[SHOC;][]{Coppejans2013} mounted on the South African Astronomical Observatory (SAAO) 1-m telescope. Observations took place on 2022 May 4, 5, and 8 and 2022 June 1, 2, 7, and 8. The observations were performed in the \textit{clear} and $g'r'i'$ filters with exposure times of either 5 or 10 s (see Table \ref{tab: observationsSALT}). The weather was clear on all nights except 2022 May 8, and we ignore these data in the rest of our analysis. 

The CCD images were reduced using the \texttt{TEA-Phot} package \citep{Bowman2019}, which was specifically designed to work with SHOC data cubes. \texttt{TEA-Phot} includes the subtraction of median combined bias images and a median combined flat-field correction to reduce the CCD images, and then uses a comparison star to produce a differential light curve of the target through adaptive elliptical aperture photometry. We used 2MASS 16542993-4337457 ($J=12.467$) as the comparison star for \texttt{TEA-Phot}. We discuss the resulting lightcurves in \S \ref{sec: OpttimingJ1654}.

\subsection{X-ray Data}
\label{sec: XrayanalysisJ1654}

\subsubsection{\textit{Swift}/XRT}
\label{sec: XRTJ1654}

We retrieved the fully calibrated \textit{Swift} data products from the NASA High Energy Astrophysics Science Archive Research Center (HEASARC) server. A log of the X-ray observations is given in Table \ref{tab: observationsJ1654}. All observations of J1654 with the X-ray Telescope \citep[XRT;][]{Burrows2005} were carried out in PC mode for a total of $10.3$ ks. We re-reduced the data using \texttt{xrtpipeline} within \texttt{HEASoft v6.29c} with \texttt{CALBD} version 20210915. We then used the \texttt{sosta} task within \texttt{ximage} to determine the source count rates, and corrected the rates for point-spread function (PSF) losses.
We extracted the source spectra from circular regions of radius 20-pixels (1 pixel $=2.36\arcsec$) and the background spectra were extracted within a nearby region of the same size. The spectra were grouped to a minimum of 1 count per bin. Using the {\it Swift}/XRT data products generator\footnote{\url{https://www.swift.ac.uk/user_objects/}} we derive an XRT enhanced position \citep{Evans2009} of RA, DEC (J2000) = $16^{h}54^m 23^{s}.40$, $-43^\circ 37\arcmin 44.0\arcsec$ with an uncertainty of $2.5\arcsec$. 


\subsubsection{\textit{Swift}/BAT}

The source does not appear in the \textit{Swift} Burst Alert Telescope \citep[BAT;][]{Barthelmy2005} catalogs \citep{Krimm2013,Oh2018,Baumgartner2013}. However, as a check, we re-analyzed data obtained on 2020 June 18 (ObsID: 03110780002) with an exposure of 3974 s in survey mode. We performed the standard BAT survey data analysis using the \texttt{HEASoft} tool (\texttt{CALDB} version 20171016), \texttt{batsurvey v6.16}, which reports the count rate at the source location in eight energy bands between $14-195$ keV. 
The analysis does not find any significant detections. The overall signal-to-noise ratio during this time period in the $14-195$ keV band is $\sim0.5$, which is consistent with a background noise fluctuation. 

To estimate the flux upper limit, we create a spectrum file using the count rate information, and generated the corresponding instrumental response files using the \texttt{HEASoft} tool, \texttt{batdrmgen}. The upper limit was determined using \texttt{XSPEC v12.12.1} assuming a power-law model with photon index $\Gamma=2$. Specifically, we adopted the \texttt{npegpow} model, which is the model used in the BAT GRB catalogs \citep{Sakamoto2011,Lien2016}. 
This analysis yields a flux upper limit of $<7.0 \times 10^{-10}$ erg cm$^{-2}$ s$^{-1}$ (90\% confidence level) in the $14-195$ keV energy range.

Furthermore, we performed a search using the BAT transient monitor \citep{Krimm2013} pipeline and found no significant detections or outbursts ($15-50$ keV) from the source over the lifetime of the \textit{Swift} mission ($\sim$18 yr).

\subsubsection{\textit{NuSTAR}}
\label{sec:  NustarJ1654}

On 2020 August 2, we carried out a \textit{NuSTAR} ToO observation of J1654 for 26 ks. We used standard tasks within the \textit{NuSTAR} Data Analysis Software pipeline (\texttt{NuSTARDAS}) within \texttt{HEASoft v6.29c} with \texttt{CALDB} version 20220131 to extract lightcurves and spectra from a circular region of size 30\arcsec{} for both FPMA and FPMB. The source region is covered by stray light, and we carefully chose the background region (60\arcsec{}) to reflect this issue. 
Spectra were grouped to a minimum of 15 counts per bin. The arrival time of photons were barycenter-corrected\footnote{\url{https://heasarc.gsfc.nasa.gov/ftools/caldb/help/barycorr.html}} to the solar system using the \textit{Chandra} source localization \citep{Anderson2014}.

\subsubsection{\textit{XMM-Newton}}
\label{sec: XMMJ1654}

We analyzed an archival \textit{XMM-Newton} observation covering the field of J1654 (Table \ref{tab: observationsJ1654}) with a total exposure of 15 ks. The observation were performed in Full Frame mode with the medium filter. We analyze and produce separate spectra for the three \textit{XMM-Newton} detectors (PN/MOS1/MOS2). 
The data were reduced and analyzed using tasks with the Science Analysis System (\texttt{SAS v18.0.0}) software using the latest CCF as of March 2022. 
We extracted the source photons in a circular region identified by the \texttt{eregionanalyse} task. The background photons were taken from an annulus surrounding the source with radii of $60\arcsec$ and $100\arcsec$, respectively. Spectra were grouped to a minimum of 1 count per bin. The response matrix and ancillary response files were obtained using the \texttt{rmfgen} and \texttt{arfgen} tasks, respectively. The event arrival times were barycenter corrected to the solar system using the \texttt{barycen} task. Using the PN data, we localize the source to RA, DEC (J2000) = $16^{h}54^m 23^{s}.43$, $-43^\circ 37\arcmin 43.4\arcsec$ with uncertainty $0.65\arcsec$ \citep[68\% CL;][]{Traulsen2020}, consistent with the \textit{Swift}/XRT position and the archival \textit{Chandra} localization.

\begin{table}
    \centering
    \caption{Photometry of the optical/infrared counterpart to J1654. The magnitudes $m_\lambda$ are reported in the AB magnitude system. We report photometry of the source in both the high- and low-states when available, as well as the difference between photometry obtained in the two states. The photometry is not corrected for interstellar reddening. Upper limits are reported at the $3\sigma$ level.
    }
    \label{tab: optcounterpartJ1654}
    \begin{tabular}{lccccc}
    \hline
    \hline
\textbf{Source} & \textbf{Filter}  & \textbf{High-state} & \textbf{Low-state}   \\
 &   & \textbf{$m_\lambda$} \textbf{(mag)} &  \textbf{$m_\lambda$} \textbf{(mag)} & \textbf{$\Delta m_\lambda$} \\
    \hline
 OM & $UVM2$ &  -- & $>21.2$ & $>1.3$ \\
 OM & $UVW1$ &  -- & $21.5\pm0.2$& 2.1 \\
 OM & $B$ &  -- & $>20.9$ & --\\
 OM & $V$ & --  &$>19.4$ &-- \\
 UVOT & $uvw2$ & $19.91\pm0.07$ & -- & -- \\
 UVOT & $uvm2$ & $19.77\pm0.11$ &  -- & $>1.3$ \\
 UVOT & $uvw1$ & $19.34\pm0.07$ &  -- & 2.1 \\
 UVOT & $u$ & $18.72\pm0.10$ & -- & -- \\
   \textit{Gaia}  & $G_{BP}$ & $18.25\pm0.16$ & -- & -- \\
  \textit{Gaia}  & $G$ & $17.83\pm0.13$ & -- & -- \\
  \textit{Gaia}  & $G_{RP}$ & $17.27\pm0.10$  & -- & -- \\
  ATLAS & $g$ & $18.36\pm0.19$ & -- & -- \\ 
  ATLAS & $r$ & $17.61\pm0.14$ & -- & -- \\ 
  ATLAS & $i$ & $17.26\pm0.13$ & -- & -- \\ 
  ATLAS & $z$ & $17.09\pm0.13$ & -- & -- \\ 
  VPHAS  & $g$  & $18.38\pm0.02$ & -- & -- \\
  VPHAS  & $r$  & $17.65\pm0.02$ & $19.94\pm0.06$ & 2.3 \\
  VPHAS & $i$ & --  & $18.97\pm0.03$ & -- \\ 
  VIRAC & $Z$ & $17.03\pm0.02$ & $18.10\pm0.05$ & 1.1 \\
  VIRAC & $Y$ & $16.67\pm0.02$ & $17.41\pm0.03$ & 0.7 \\
  VIRAC & $J$  & $16.45\pm0.03$  & $16.66\pm0.05$ & 0.2 \\
  VIRAC & $H$ & $16.50\pm0.05$  & $16.66\pm0.08$ & 0.16 \\
  VIRAC & $K$ & $16.80\pm0.03$ & -- & $<0.1$ \\
 GLIMPSE & $3.6\mu$m & $16.84\pm0.13$ & -- &--  \\
 GLIMPSE & $4.5\mu$m & $16.58\pm0.16$ & -- &--  \\
    \hline
    \hline
    \end{tabular}
\end{table}

\section{Results} 
\label{sec: resultsJ1654}

\begin{figure*}
\centering
\includegraphics[width=1\columnwidth]{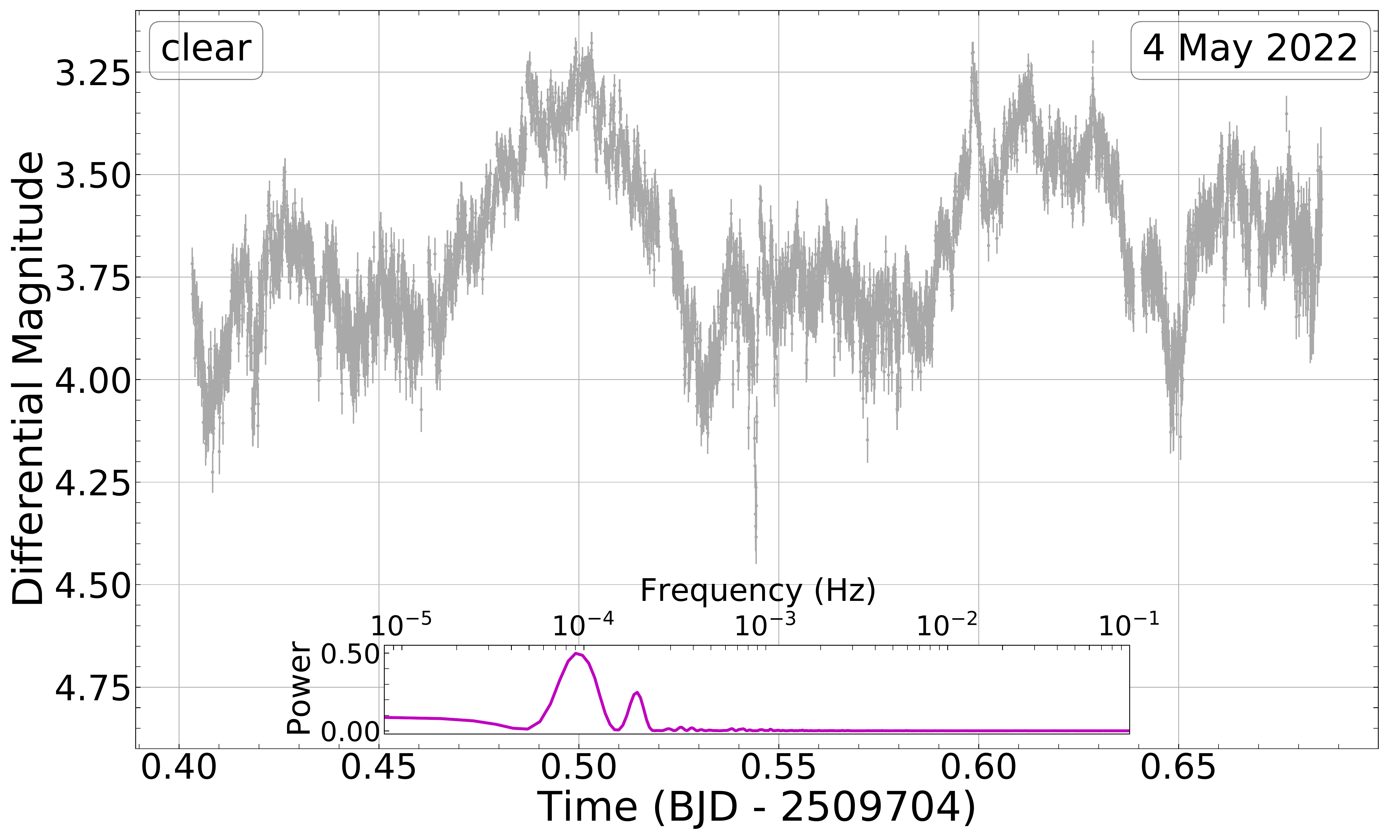}
\includegraphics[width=1\columnwidth]{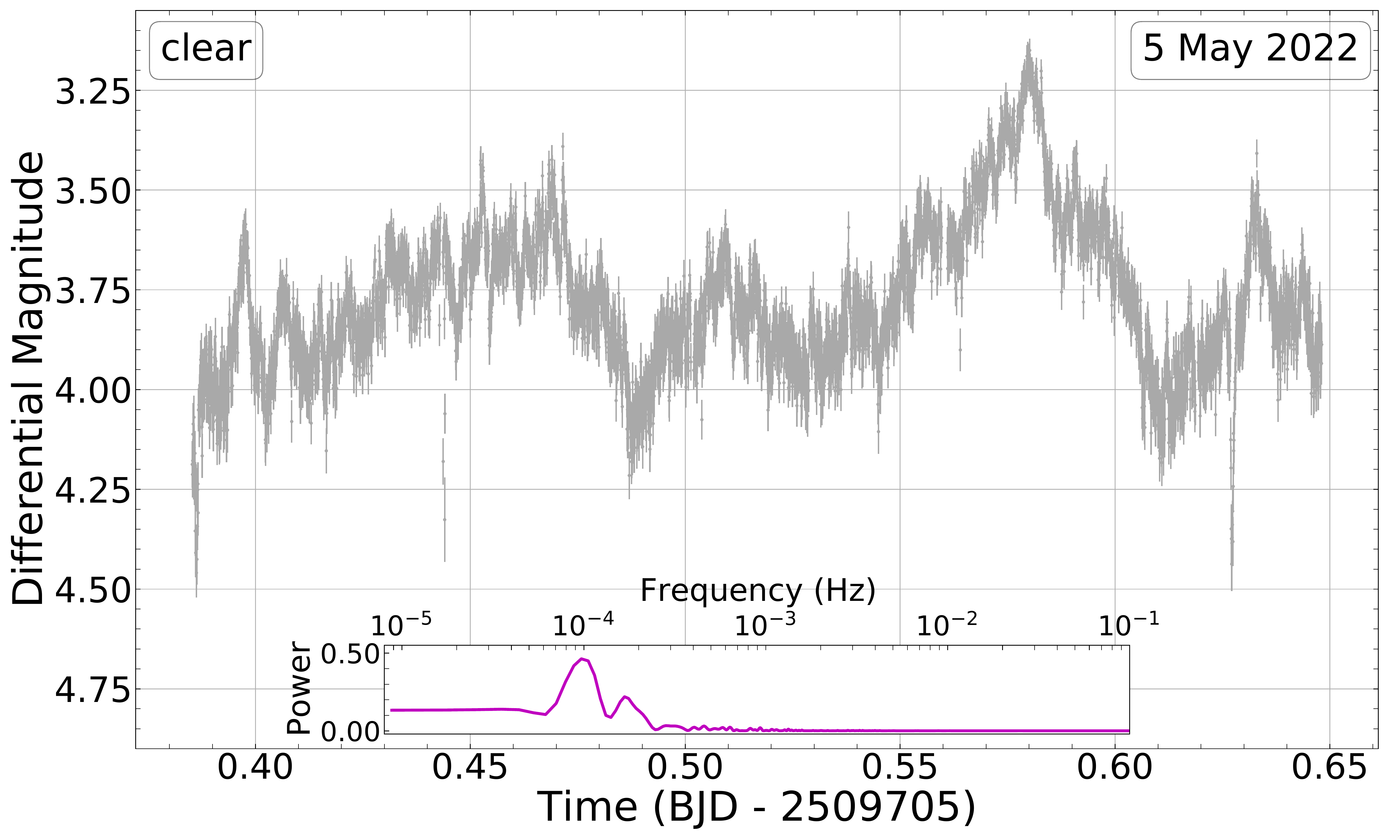}
\includegraphics[width=1\columnwidth]{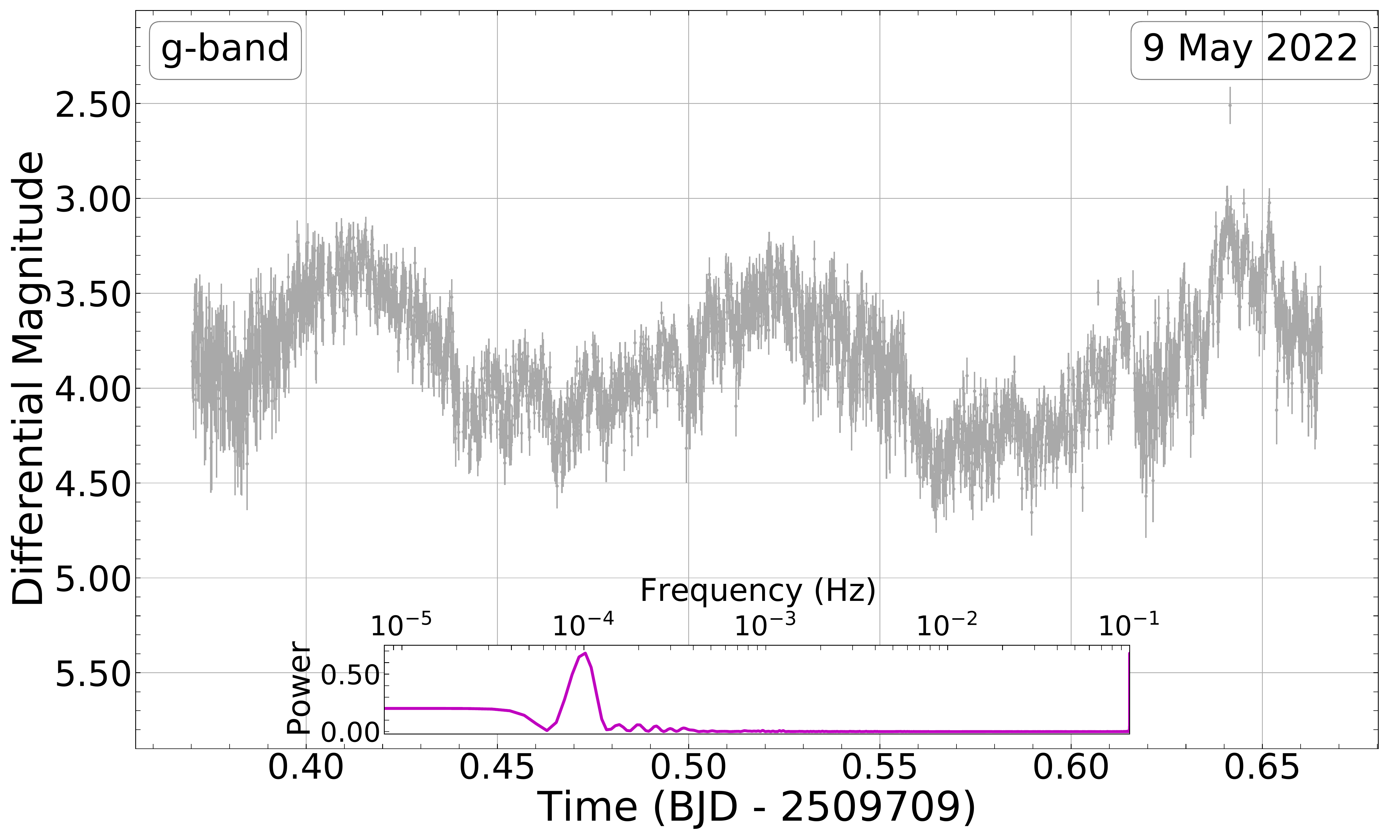}
\includegraphics[width=1\columnwidth]{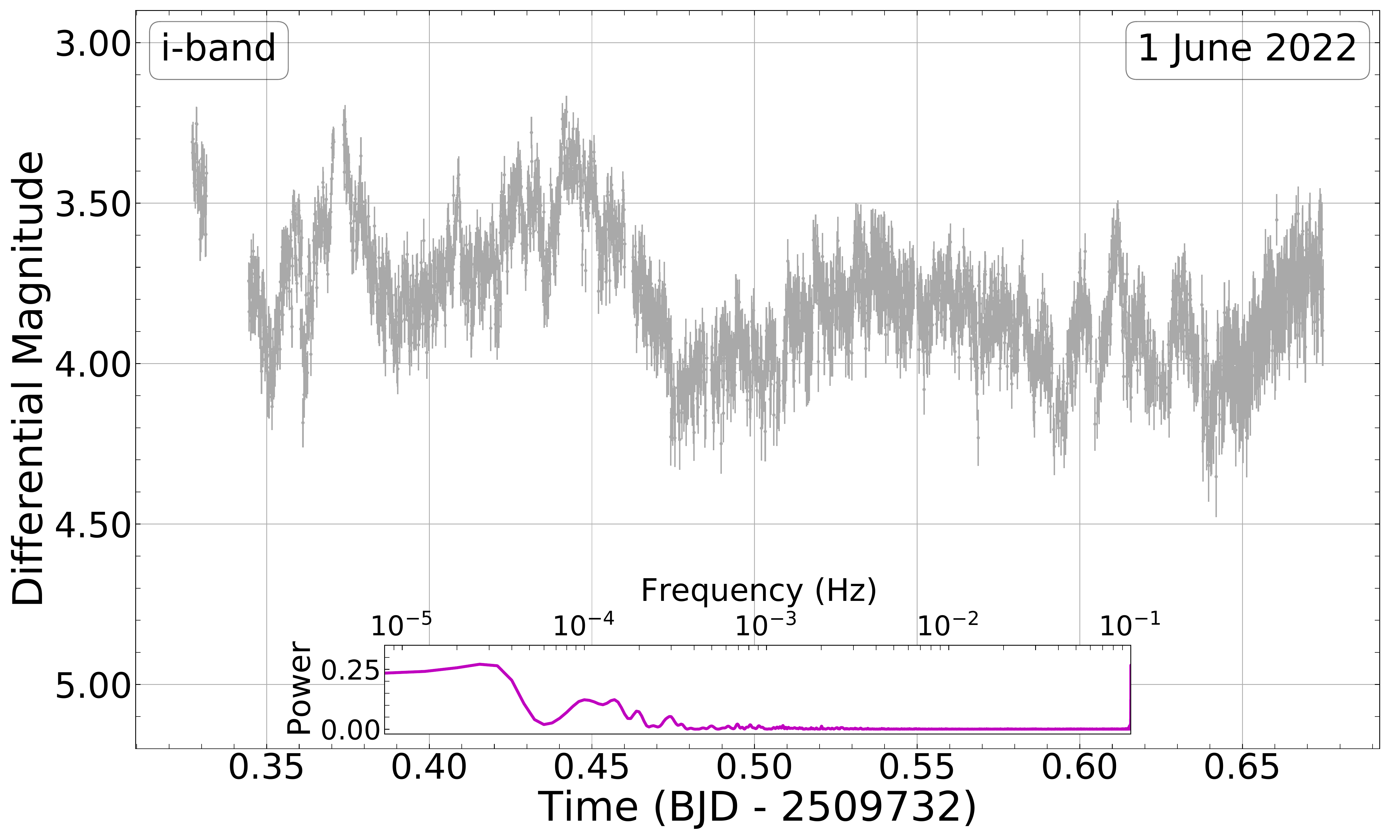}
\includegraphics[width=1\columnwidth]{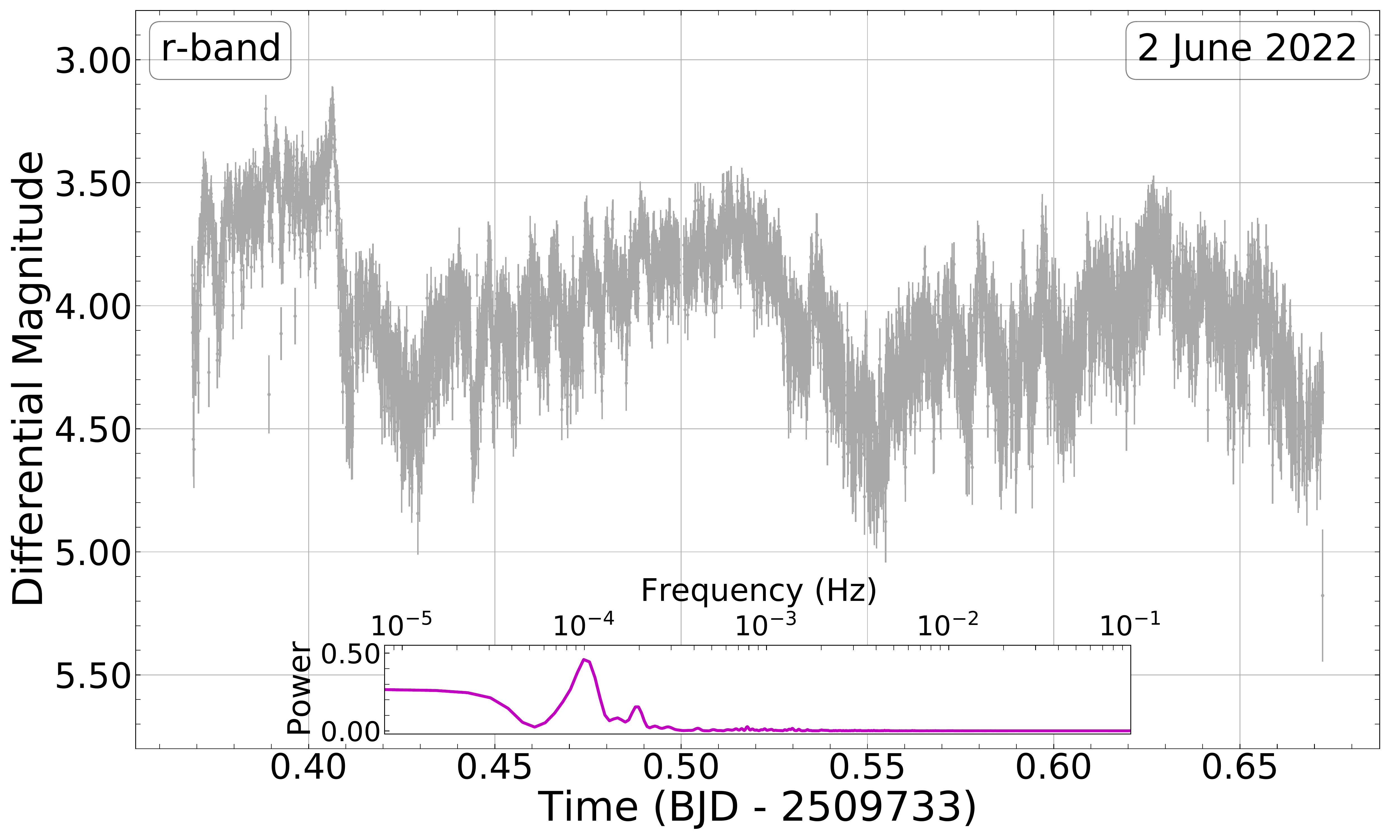}
\includegraphics[width=1\columnwidth]{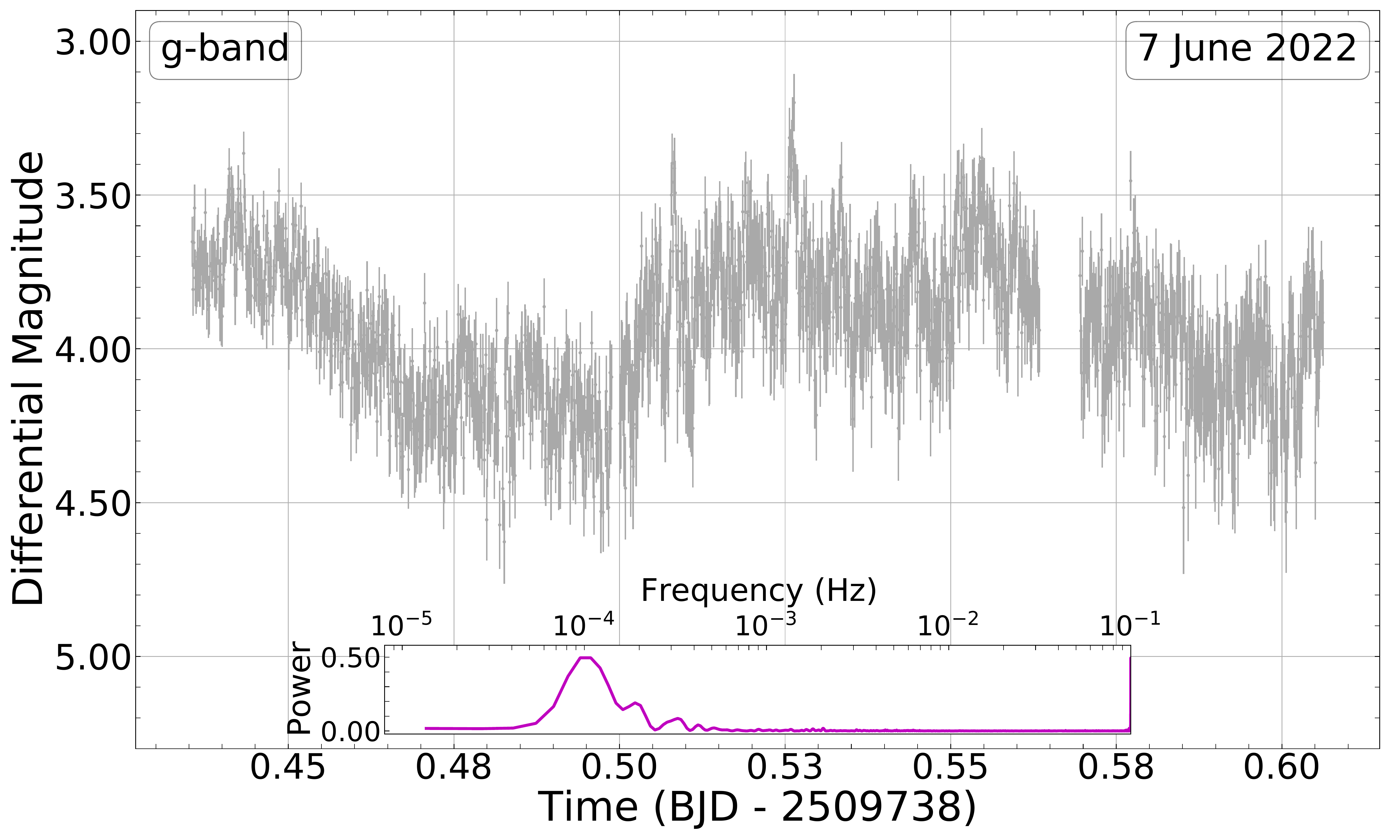}
\caption{Optical lightcurves of J1654 from the 1-m SAAO telescope obtained on different nights, labeled by their Barycentric Julian Date (BJD). The photometry is measured as differential instrumental magnitudes, dependent on the observing conditions on each night.  The inset displays the periodogram for each night. 
}
\label{fig: optical_lightcurves}
\end{figure*}

\begin{figure}
\centering
\includegraphics[width=1\columnwidth]{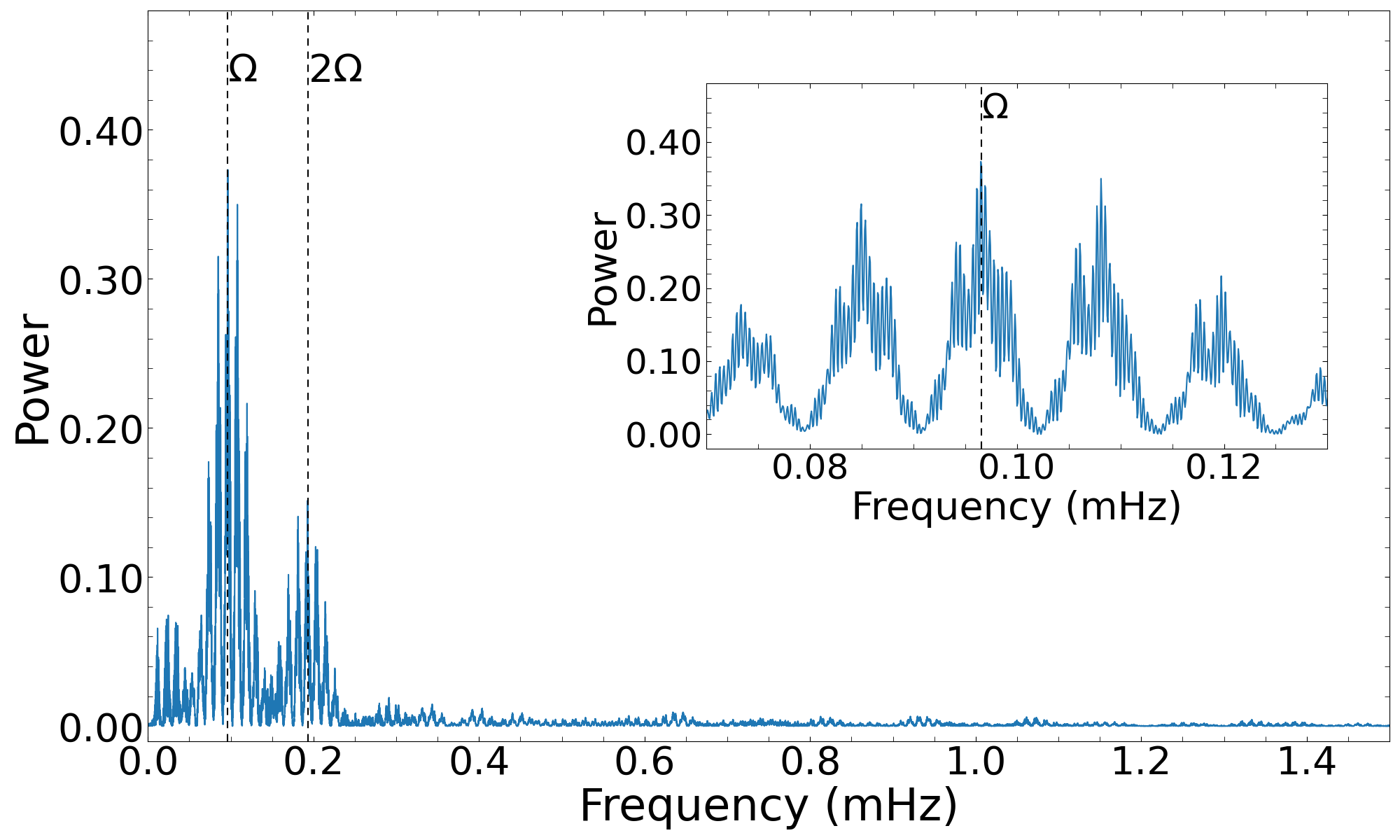}
\caption{Lomb–Scargle periodogram of the combined optical data obtained with the SAAO. The orbital period $\Omega$ (main peak) and its harmonic are observed. The inset shows a zoom-in on the frequency corresponding to the orbital period of 2.87 hr, and its aliases.}   
\label{fig: optical_periodogram}
\end{figure}

\subsection{Optical Lightcurves and Timing Analysis}
\label{sec: OpttimingJ1654}


Our initial high-speed photometry lightcurves exhibited characteristic features of polars, such as dips, flares, and flickering, with amplitudes of up to $1$ mag. 
In order to search for an orbital period, 
we employed a Lomb-Scargle period search with \texttt{Gatspy} \citep{Gatspy} to identify significant coherent pulsations in the differential lightcurves produced by \texttt{TEA-Phot} (Figure \ref{fig: optical_lightcurves}). We built periodograms for each individual night (Figure \ref{fig: optical_lightcurves}) as well as a combined periodogram after normalizing the lightcurves to the same scale (Figure \ref{fig: optical_periodogram}). We identify a significant signal at a period of 2.87 hr ($10359\pm111$ s), which we interpret as the orbital period. The significance of this signal is at the $>5\sigma$ level. 
We also identify a harmonic of this period. No other significant lower frequency signals were identified in the periodogram at expected beat frequencies for IP systems. 

We note that an independent analysis of data from the Transiting Exoplanet Survey Satellite was carried out by \citet{Takata2022} (their source G596) which finds a period of 0.12 d (2.88 hr). Our observations are in agreement with their result. 

\subsection{X-ray Lightcurves and Timing Analysis}
\label{sec: XraytimingJ1654}

\begin{figure}
\centering
\includegraphics[width=1\columnwidth]{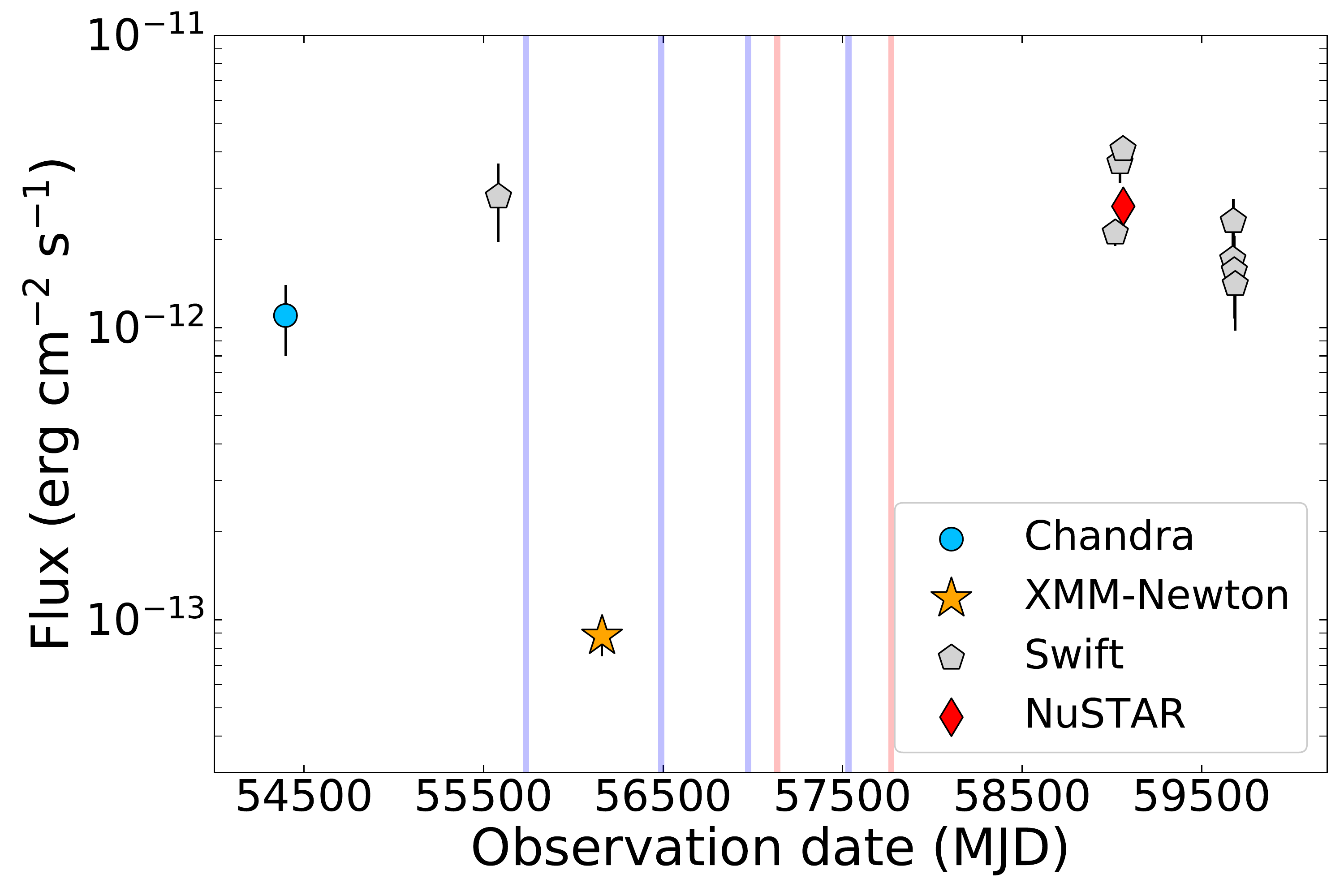}
\includegraphics[width=1\columnwidth]{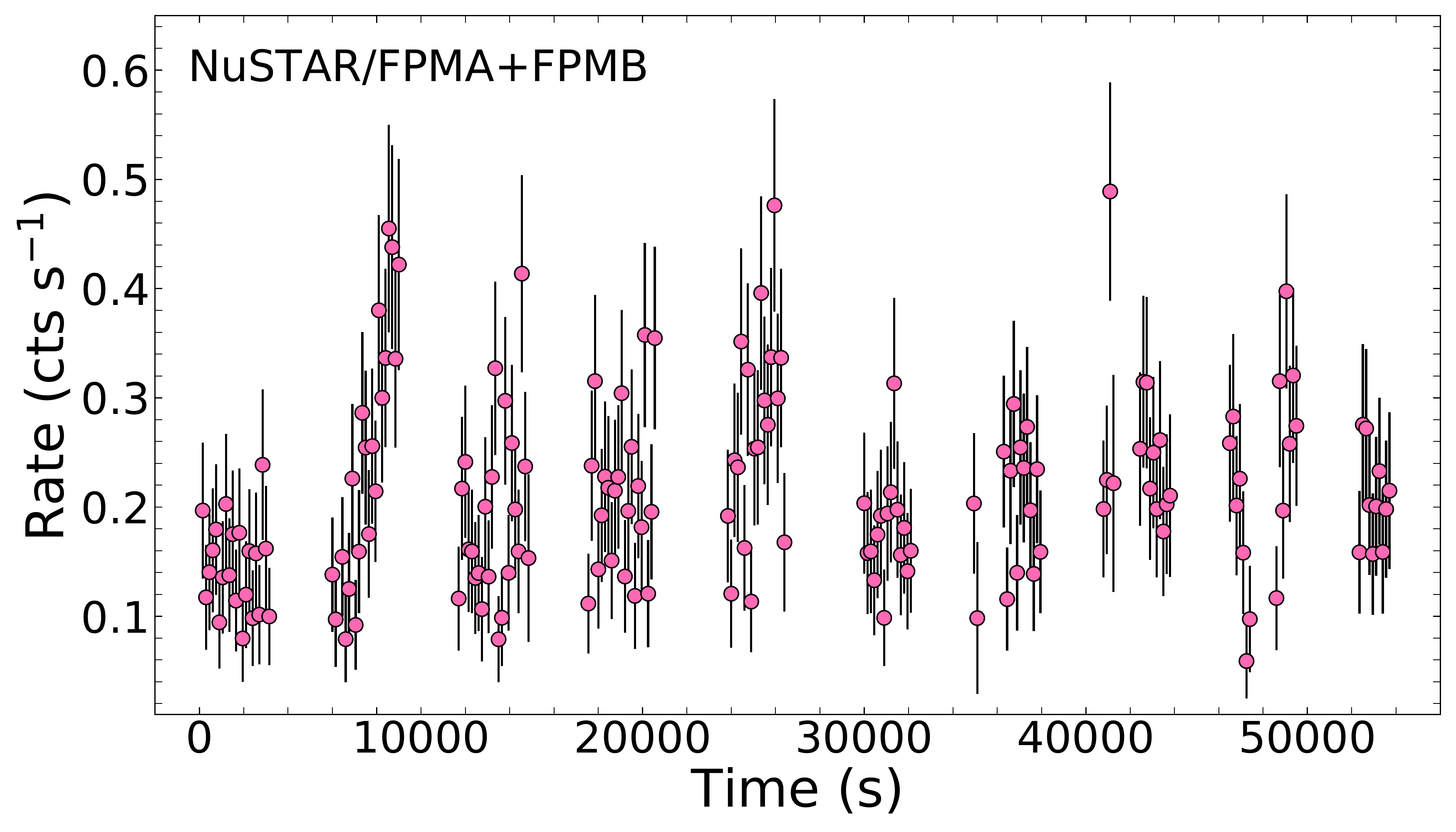}
\caption{
\textbf{Top:} Long-term lightcurve ($0.3-10$ keV) of all X-ray observations included in this work. Vertical lines mark known times of high- (blue) and low-states (red) based on archival optical/infrared observations. 
\textbf{Bottom:} Background subtracted \textit{NuSTAR} lightcurve of J1654 in the $3-79$ keV energy range. We combined the count rates from both FPMA and FPMB. The time bin is 150 s.
}
\label{fig: J1654_nustar_lc}
\end{figure}

We compiled available archival X-ray observations of J1654 (Table \ref{tab: observationsJ1654}) in order to produce a long term X-ray lightcurve of the source. We display this lightcurve, covering a time period from 2007 to 2022, in Figure \ref{fig: J1654_nustar_lc} (top). Based on our analysis, we identify that the \textit{XMM-Newton} observation from 2012 August 20 was obtained during a low-state \citep{Warner1999}, while all other available X-ray observations, including \textit{ROSAT} (1990) and \textit{ASCA}, detected the source during a high-state. We discuss a comparison between the \textit{NuSTAR} high-state and \textit{XMM-Newton} low-state in \S \ref{sec: XrayspecanalysisJ1654}.

We searched the \textit{NuSTAR} and \textit{XMM-Newton} observations for coherent pulsations using both Z$^{2}$ statistics \citep{Buccheri1983} and a Lomb-Scargle frequency analysis \citep{Scargle1982}. We searched both the $3-10$ keV and $3-79$ keV \textit{NuSTAR} lightcurve (Figure \ref{fig: J1654_nustar_lc}; bottom), and the $0.3-3$ keV and $0.3-10$ keV \textit{XMM-Newton} lightcurve. For the \textit{NuSTAR} data we searched both the FMPA and FMPB data separately, as well as the combined events file. We likewise searched both PN, MOS1, MOS2, and the combined events files for \textit{XMM-Newton} for both energy ranges. No significant signal was identified in our analysis. While the \textit{NuSTAR} observation does indeed cover about 5.3 orbital cycles, we did not find any significant signal at 2.87 hr in either dataset, nor at any other frequency. We are able to set a $3\sigma$ upper limit to the RMS pulsed fraction of $<9\%$ in the $3-79$ keV energy range using the \textit{NuSTAR} data (FPMA/FPMB) and $<25\%$ in the $0.3-10$ keV energy range using the \textit{XMM-Newton} data (PN/MOS1/MOS2). 

The lack of other signals in the data besides the orbital period at 2.87 hr identified in our optical data (\S \ref{sec: OpttimingJ1654}) is in agreement with the independent analysis of \citet{Takata2022}. \citet{Takata2022} likewise did not identify any periods at shorter or longer values than the orbit, which would have been expected for an IP system.

\subsection{Spectral Energy Distribution}
\label{sec: sedsection}

\begin{figure}
\centering
\includegraphics[width=1\columnwidth]{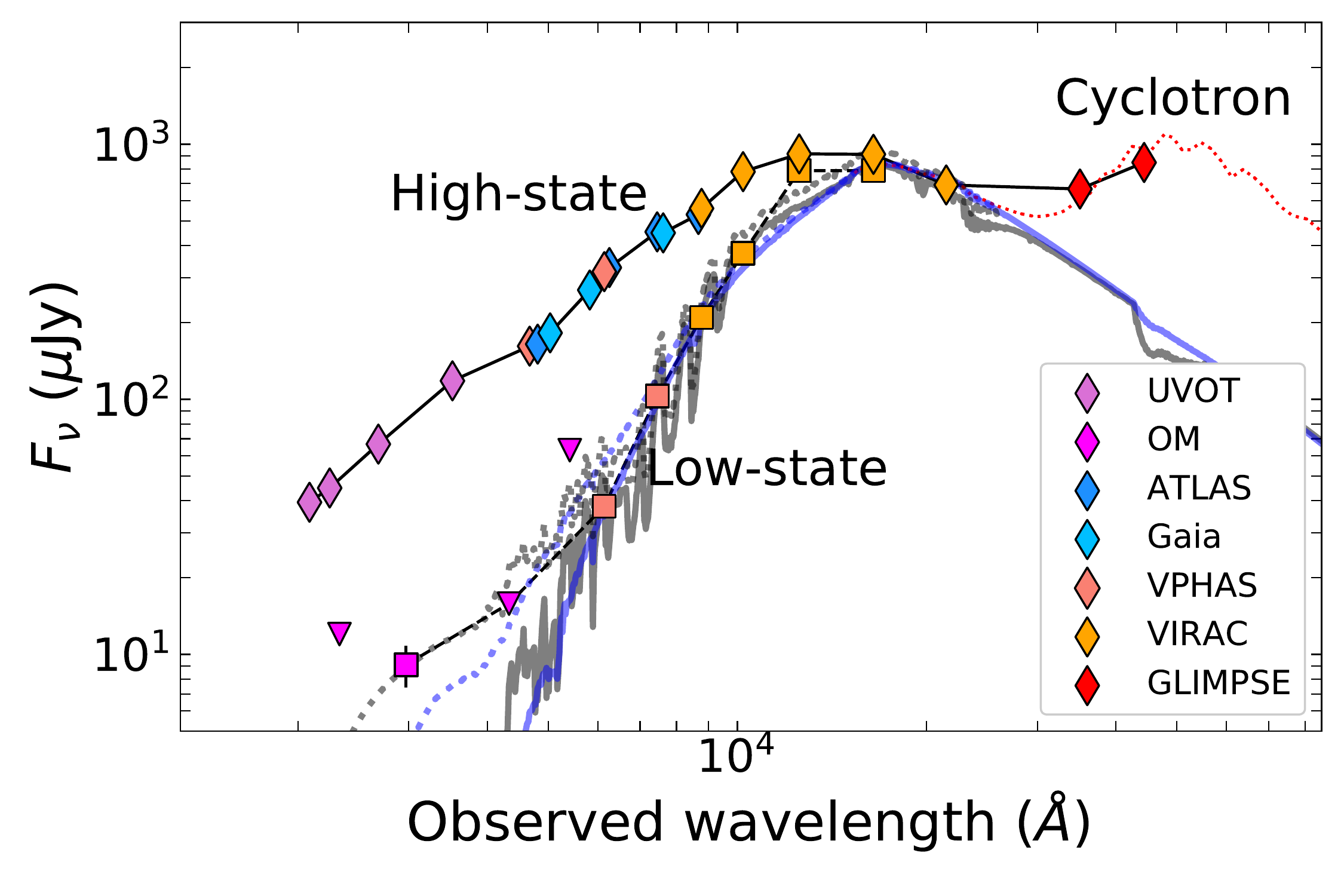}
\caption{
SED of the ultraviolet, optical, and infrared counterpart to J1654. The solid black line connects photometry (diamonds) obtained during the high-state, whereas the dashed black line (and squares) shows the low-state. The data are not corrected for extinction; downward triangles represent $3\sigma$ upper limits. 
The thick, solid blue line shows the best fit stellar model from \texttt{ARIADNE} (K dwarf), and the solid gray line is a less extinct M dwarf (see \S \ref{sec: sedsection}). 
The dotted blue line represents the addition of a WD with $T=120,000$ K and radius $R=0.01R_\odot$, modeled as a simple blackbody, extinct by $A_V\sim4.9$ mag. The dotted gray line represents a WD with $T=12,000$ extinct by $A_V\sim0.5$ mag. 
The red dashed line is the addition of cyclotron emission from a polar with a magnetic field strength of $3.5$ MG. 
}
\label{fig: sedJ1654}
\end{figure}

We compiled all available ultraviolet, optical, and infrared  photometry of J1654 (Table \ref{tab: optcounterpartJ1654}) in order to build the source spectral energy distribution (SED). We discovered significant variability in the UV and optical filters by $\sim1-2$ mag,  with the largest variability amplitude in the UV. The variability was identified as the transition between a high- and low-state, as shown in Figure \ref{fig: sedJ1654}. In the high-state, the UV/optical emission from the accretion column dominates, but during periods of low accretion we observe the contribution of the secondary star, peaking in the near-infrared. An additional component, interpreted as cyclotron emission, is observed at longer wavelengths ($3-5\mu$m from \textit{Spitzer}/GLIMPSE). At UV wavelengths, we begin to observe either the contribution from the WD, modeled as a simple blackbody, or emission from the accretion column that is fainter due to a lower accretion rate during those periods. Given the sparse UV data during the low-state it is not possible to disentangle between those contributions. Further UV observations would be critical to constrain these possibilities.

The low-state SED is well sampled in the optical and infrared (Table \ref{tab: optcounterpartJ1654}), allowing us to identify the spectral type of the secondary star. We applied the \texttt{ARIADNE} \citep{ariadne} code to perform dynamic nested sampling with \texttt{dynesty} \citep{dynesty} in order to determine the best fit secondary star parameters over a range of stellar models \citep[e.g.,][]{Kurucz1993,Castelli2004,Husser2013}. We find an effective temperature of $4700^{+800}_{-200}$ K, surface gravity $\log g = 4.7^{+0.6}_{-0.4}$, stellar radius $R=0.32^{+0.07}_{-0.04} R_\odot$, metallicity [Fe/H]$=-0.12\pm0.15$, a significant extinction $A_V=4.9^{+1.1}_{-0.7}$ mag, and distance of $454\pm20$ pc, which is in good agreement with the \textit{Gaia} estimate \citep{Bailer-Jones2018,Bailer-Jones2021}. A corner plot of these results is displayed in Figure \ref{fig: ariadne} and the best fit is shown in Figure \ref{fig: sedJ1654} (blue solid line). Overall this is consistent with a star of spectral type K3.5V.

The orbital period and $i-K$ color also favor a main-sequence donor star of late spectral type, but instead closer to spectral types between M3V to M5V \citep{Knigge2006}. 
Without optical spectroscopy of the secondary star during the low-state we cannot rule out other spectral types, but note the SED clearly favors a late spectral type.  
Moreover, using a photometric distance estimate based on the $K$-band apparent magnitude \citep{Bailey1981,Warner1987,Knigge2011}, we find a lower limit to the distance of $d\gtrsim300$ pc, which is consistent with the \textit{Gaia} result \citep{Bailer-Jones2021}. The estimate is a lower limit given that we cannot rule out an additional emission component (e.g., a circumbinary dust disk or cyclotron radiation; \citealt{Brinkworth2007,Harrison2015}), besides the secondary, contributing to the observed $K$-band flux.

We note that the \texttt{ARIADNE} model only represents one possible range of solutions. For example, the $H$-band photometric point (Figure \ref{fig: sedJ1654}) is underestimated in this model, though this could also be due to the source not fully being in the low-state at the time of this measurement. As an additional check on the secondary star spectral type, we test whether a less extinct M dwarf can explain the low-state SED. We performed a ``by-eye'' check, comparing Kurucz models \citep{Kurucz1993} to the SED. We find an appropriate match for a M2V dwarf at a distance of 460 pc, with temperature $T\sim3500$ K, surface gravity $\log g \sim 4.7$, metallicity $\log Z/Z_\odot \sim 1$, and radius $R\sim0.3 R_\odot$ \citep[see, e.g.,][]{Parsons2018} that is extinct by $A_V\sim0.5$ mag. This model is also shown in Figure \ref{fig: sedJ1654} (gray solid line). With the available constraints to the spectral type, we cannot exclude either of the two scenarios (e.g., K vs M dwarf), but can conclude that a late-spectral type is the most likely.

In order to determine the contribution from cyclotron radiation in the infrared, and in particular at \textit{Spitzer} wavelengths, we applied the cyclotron models from \citet{Potter2002}. These calculations utilize the shock structures from \citet{Cropper1999} and the cyclotron opacity and radiative transfer from \citet{Meggitt1982,Wickramasinghe1985}. The model is described by the WD mass, accretion rate, and magnetic field strength. The detailed morphology of the cyclotron emission, in particular the cyclotron humps, will depend on the WD mass and accretion rate. However, the wavelength at which the cyclotron emission peaks is dominated by the magnetic field strength, with less influence from the WD mass and accretion rate. In the case of J1654, we are limited by the fact that we do not detect phase-dependent cyclotron humps in our optical spectroscopy, which would be required to obtain robust limits on the WD mass and accretion rate. Instead, we simply re-scale the model (for a specific $B$ field) to match the peak of the excess IR emission in Figure \ref{fig: sedJ1654}.

We find that if the infrared excess is due to cyclotron radiation a low magnetic field strength of $B\lesssim3.5$ MG is required. Any larger magnetic field  would push the excess towards the optical. We have simply shown that if the main cause of the IR emission is cyclotron, that the magnetic field strength must be low to match the IR peak. We cannot, however, rule out the contribution of a circumbinary disk at these wavelengths \citep{Brinkworth2007,Harrison2015}. A circumbinary disk may reasonably explain the large extinction inferred from the SED. In either case, the lack of cyclotron humps in our phase-dependent optical spectra imply that small magnetic field strength  would be consistent with the excess from \textit{Spitzer}. Despite the uncertainty in the spectral type of the secondary star, due to the similarity of the models (Figure \ref{fig: sedJ1654}), it does not change our estimate of the magnetic field.

We further note that the detection of the source in the $UVW1$ filter by the \textit{XMM-Newton}/OM during the low-state may indeed be the contribution of the WD. We consider this in both of the scenarios for the secondary star spectral type. In Figure \ref{fig: sedJ1654}, the dotted blue line shows the contribution of a thermal blackbody with $T=120,000$ K at $d=460$ pc added on top of the K dwarf model that has a large extinction $(A_V\sim4.9$ mag).
The high temperature is a direct result of the large extinction implied by the secondary star SED. In the M dwarf scenario, which implies a low extinction ($A_V\sim0.5$ mag), we find a temperature of $T=12,000$ K. These scenarios are quite different, and without further UV observations it is difficult to favor one over the other. An additional possibility is that the $UVW1$ detection during the low-state is still due to the accretion disk, and simply fainter due to a lower accretion rate.



\subsection{Optical Spectroscopy}
\label{sec: OptspecanalysisJ1654}

 \begin{figure*}
\centering
\includegraphics[width=2\columnwidth]{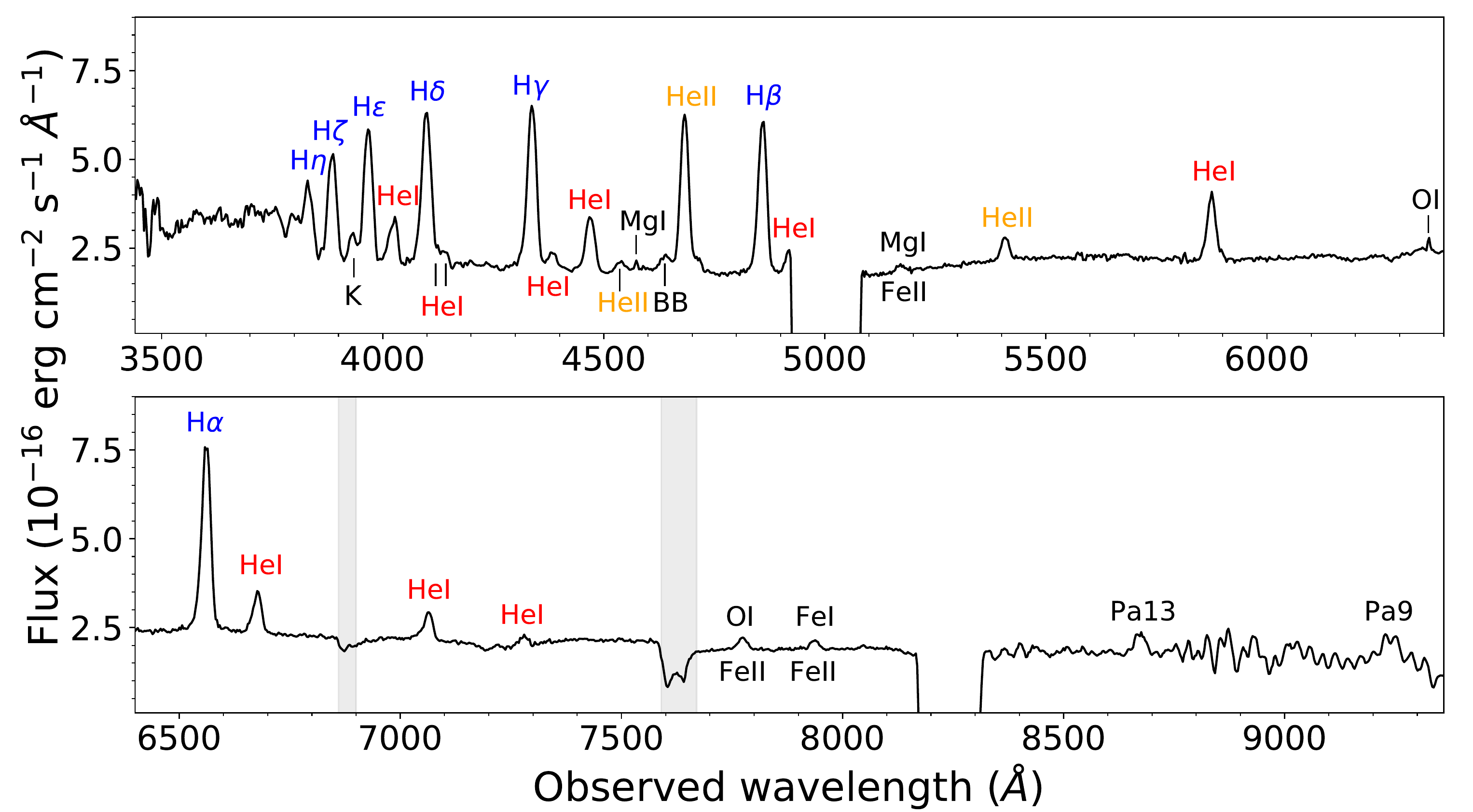}
\caption{Mean spectrum of J1654 obtained with SALT. We have labeled the primary emission features: Balmer lines in blue, HeI lines in red, HeII lines in orange, and other features in black, including the Bowen Blend (BB; NIII/CIII). Chips gaps are located at 5000 \AA\, and 8250. Gray vertical regions mark telluric features. 
}
\label{fig: spectraJ1654}
\end{figure*}



The initial optical spectrum of J1654 obtained with SALT (\S \ref{sec: saltJ1654}) is displayed in Figure \ref{fig: spectraJ1654}. 
We detected a flat continuum with a slight blue slope, peppered with emission features. 
There are no obvious features related to the secondary star in the mean spectrum. 
In total, we identify 29 high significance emission lines, including hydrogen lines of the Balmer and Paschen series, HeI ($\lambda$3487,	3889, 4026, 4144, 4387, 4471, 4921, 5875, 6678, 7065, 7281), HeII ($\lambda$4511, 4686, 5411), CaII K, MgI, OI, and the $\lambda4640$ Bowen blend (a CIII+NIII line complex). The emission features are narrow ($<1000$ km s$^{-1}$). This suggests emission from a hot spot, as opposed to the velocity-broadened inner regions of an accretion disk. 

The wealth of helium lines, in particular HeII $\lambda4511$, $\lambda4686$, and $\lambda5411$, are characteristic of mCVs. We note that further HeII lines (e.g., $\lambda$4101, 4340, 4861, 6562) are likely overlapping with the Balmer lines. 

We use the observed line fluxes (Table \ref{tab: linesJ1654}) to compute the Balmer decrements. The line fluxes were determined by modeling each line as single Gaussian on top of the continuum emission. 
The ratios H$\alpha$/H$\beta\approx1.36$ and H$\gamma$/H$\beta\approx1.16$ have not been corrected for the uncertain Galactic extinction, but still signify an inverted Balmer decrement. The inverse Balmer decrement  is in excess to expectation based on optically thin ``case B'' recombination \citep{Pottasch1984}, and is, therefore, indicative of emission from an optically thick accretion region \citep{Stockman1977,Williams1980,Schwope1997}, typical of CVs. We further identify the observed ratio of H$\beta/$HeII $\lambda4686\approx1.0$ as evidence that J1654 is a polar  \citep{Williams1982,Williams1989,Silber1992,Warner2003}. 

If we assume $A_V\approx 4.9$ mag as derived in \S \ref{sec: sedsection}, we can correct the line fluxes using a \citet{Cardelli1989} Milky Way extinction law. This yields H$\alpha$/H$\beta\approx 0.3$, H$\gamma$/H$\beta\approx2.5$, and H$\beta/$HeII $\lambda4686\approx 0.76$. However, $A_V\approx 4.9$ mag is well above the $A_V\approx 0.13-0.35$ mag implied by Galactic dust maps over this distance range \citep{Amores2005,Amores2021}, which would yield ratios of order unity. We further note that the spectra do not appear significantly reddened nor are there significant NaI absorption lines, which would suggest a lower extinction than $A_V\approx4.9$ mag. The order unity line ratios, when not de-reddened, further favor a lower extinction value.

\begin{figure}
\centering
\includegraphics[width=\columnwidth]{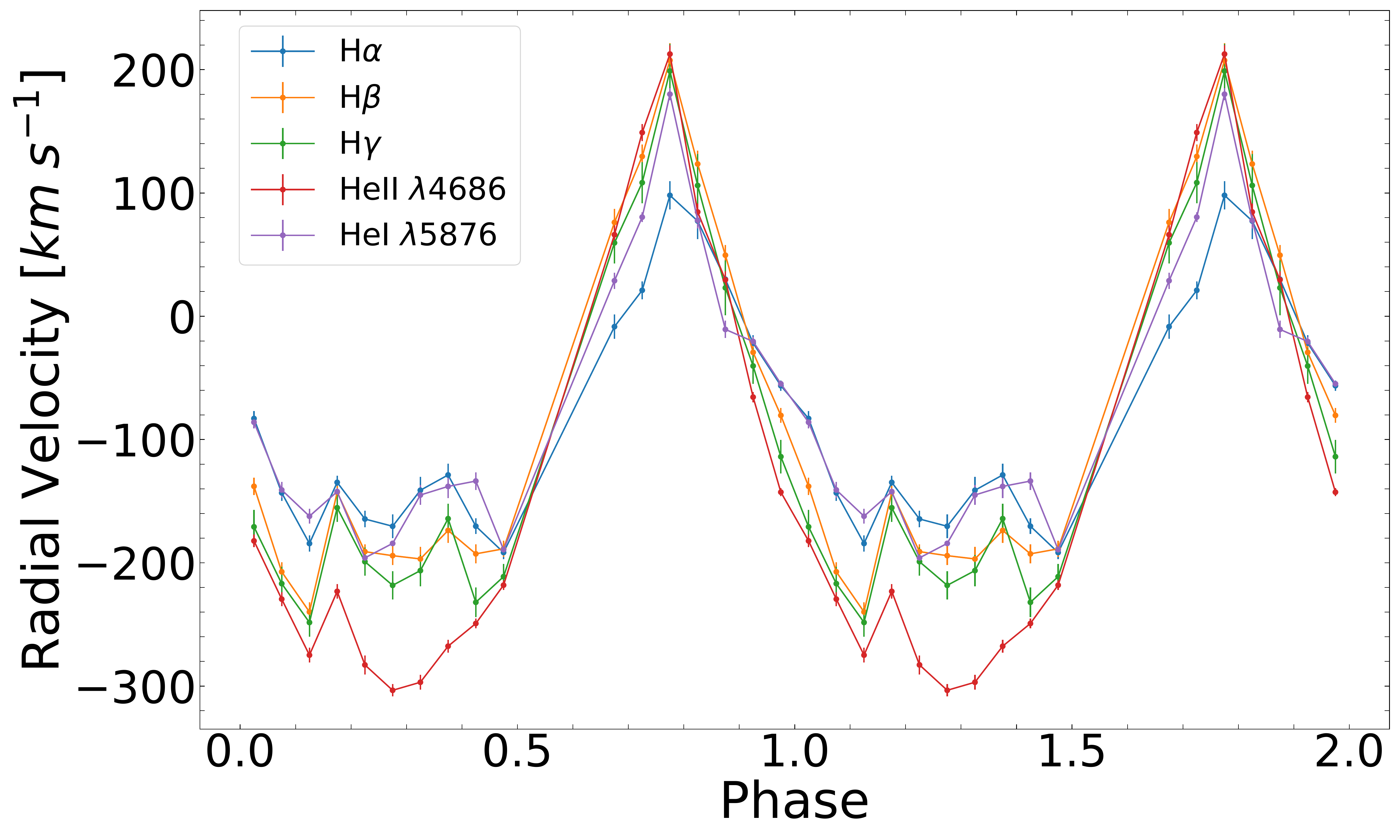}
\caption{Radial velocity versus orbital phase for the H$\alpha$, H$\beta$, H$\gamma$, HeII $\lambda4686$, and HeI $\lambda5876$ emission lines.
}
\label{fig: RV}
\end{figure}

Using the frame-transfer spectroscopy, we measured the radial velocities of H$\alpha$, H$\beta$, H$\gamma$, HeII $\lambda4686$, and HeI $\lambda5876$ by fitting Gaussian functions using the \texttt{lmfit} package \citep{lmfit} to derive the central wavelength, amplitude, and FWHM. 
The radial velocities of these lines as a function of phase are shown in Figure \ref{fig: RV}. The HeI $\lambda5876$ and HeII $\lambda4686$ lines display a qualitatively similar behavior to the Balmer lines. This likely implies that the emission region is the same.


 \begin{table}
    \centering
    \caption{Spectral features observed in the mean SALT spectrum from 26 April 2022. The line fluxes have not been corrected for extinction. 
    }
    \label{tab: linesJ1654}
    \begin{tabular}{lcc}
    \hline
    \hline
\textbf{Feature}  & \textbf{FWHM} &  \textbf{Flux} \\
 & \textbf{(\AA)} &   \textbf{(erg cm$^{-2}$ s$^{-1}$)} \\
    \hline
  H$\alpha$  & 23.0 & $1.3\times10^{-14}$ \\
  H$\beta$  & 21.8  & $9.5\times10^{-15}$ \\
  H$\gamma$  & 22.8 & $1.1\times10^{-14}$ \\
  HeII $\lambda4686$  & 20.6 &$9.5\times10^{-15}$ \\
  HeI $\lambda5876$ & 22.5 & $4.3\times10^{-15}$ \\
    \hline
    \hline
    \end{tabular}
\end{table}


\subsubsection{X-ray Spectral Analysis}
\label{sec: XrayspecanalysisJ1654}



We performed a time-averaged spectral analysis of the \textit{Swift}/XRT, \textit{XMM-Newton} (MOS1/MOS2/PN), and \textit{NuSTAR} (FPMA/FPMB) spectra using \texttt{XSPEC v12.12.0} \citep{Arnaud1996}. We applied the ISM abundance table from \citet{Wilms2000} and the photoelectric absorption cross-sections presented by  \citet{Verner1996}. Spectral fits were performed in the $3-20$ keV energy range for \textit{NuSTAR} (the background dominates above 20 keV) and $0.3-10$ keV for \textit{Swift} and \textit{XMM-Newton}. We determined the best fit parameters by minimizing the Cash statistic \citep{Cash1979}.

We began by modeling the time-averaged PC mode spectrum for all XRT observations ($10.3$ ks). Individual shorter exposures yield large model uncertainties, and the time-averaged spectrum allows us to better constrain the parameters. The best fit absorbed power-law model (\texttt{con*tbabs*pow}) has $N_H=(2.4\pm1.4)\times 10^{21}$ cm$^{-2}$ and $\Gamma=1.46\pm0.25$. This yields a time-averaged unabsorbed flux $F_X=(2.56\pm0.30)\times 10^{-12}$ erg cm$^{-2}$ s$^{-1}$ in the $0.3-10$ keV energy band. For these parameters, we derive an energy conversion factor (ECF) of $5.9\times 10^{-11}$ erg cm$^{-2}$ cts$^{-1}$, which we use to convert the count rates in individual exposures to an unabsorbed flux (Figure \ref{fig: J1654_nustar_lc}; Top Panel).

\begin{figure}
\centering
\includegraphics[width=1\columnwidth]{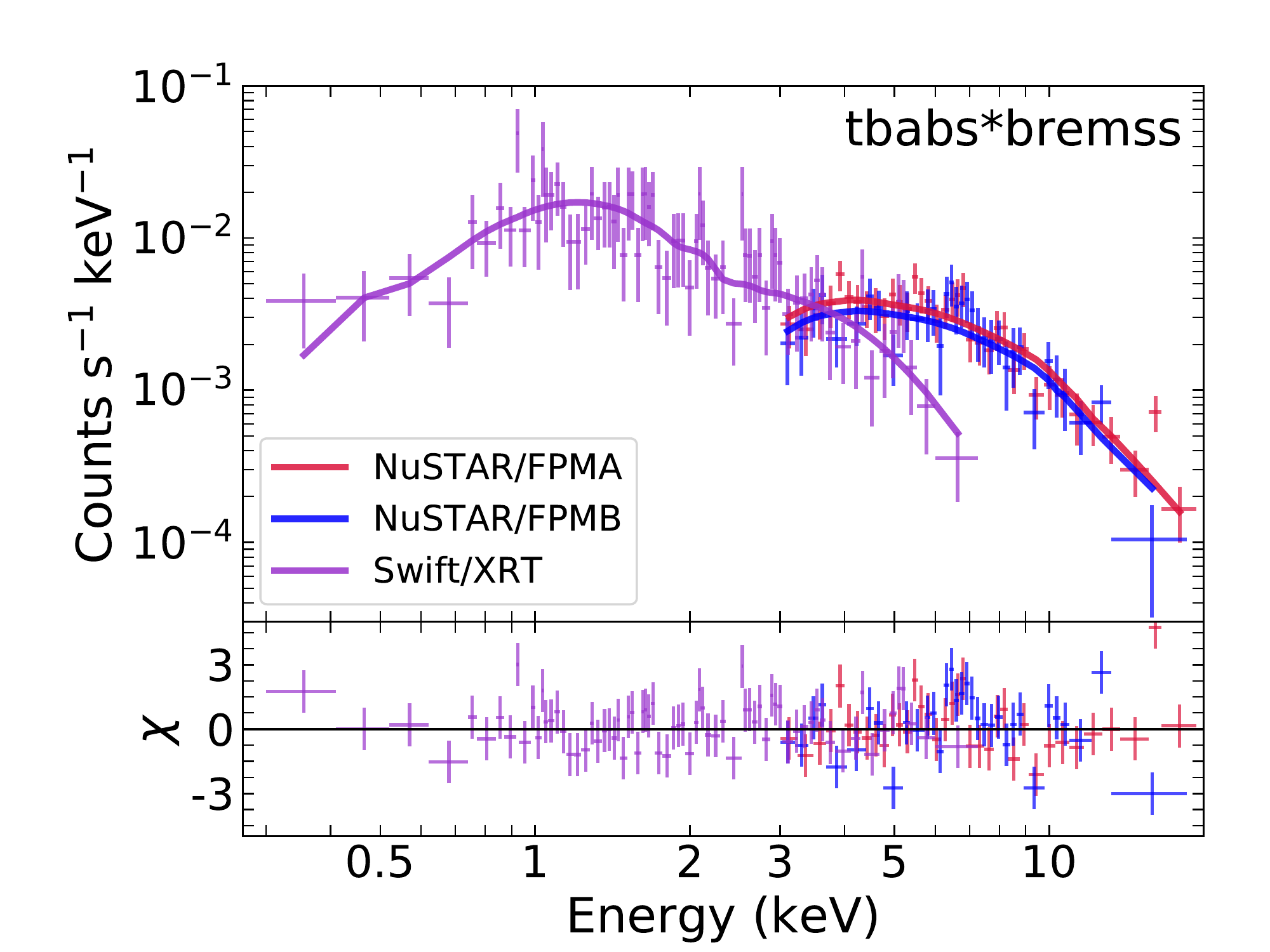}
\caption{X-ray spectra of J1654 in the high-state from \textit{Swift} and \textit{NuSTAR}. The best-fit model (\texttt{con*tbabs*bremss}) is shown. The spectra have been re-binned for display purposes. There is a hint of a Fe K$\alpha$ line at $6.7$ keV in the \textit{NuSTAR} spectra. 
 }
\label{fig: x-ray_spectrumJ1654}
\end{figure}

As all of the \textit{Swift} observations occur within the high-state, we performed a joint fit with the \textit{NuSTAR} data, which was also obtained during the high-state (Figure \ref{fig: J1654_nustar_lc}; Top Panel). 
Based on the source's classification as a mCV \citep{Takata2022}, we tested three physically motivated models: \textit{i}) thermal bremsstrahlung radiation (\texttt{constant$*$tbabs$*$bremss}), likely caused as the magnetically funnelled accretion stream shocks above the WD poles, \textit{ii}) a cooling flow model (\texttt{constant$*$tbabs$*$mkcflow}) that self-consistently accounts for both bremsstrahlung radiation and line cooling \citep{Mushotzky1988}, and, lastly, \textit{iii}) a model for bremsstrahlung radiation in the post-shock structures (PSRs) of magnetized WDs (\texttt{constant$*$tbabs$*$ipolar}; \citealt{Suleimanov2016,Suleimanov2019}).  
This last model is parametrized by the WD mass $M_\textrm{WD}$ and relative magnetospheric radius $R_m$ (hereafter referred to as the fall-height). Following \citet{Shaw2020}, we froze the fall-height to $R_m=1000 R_\textrm{WD}$. 
In the \texttt{mkcflow} model, we froze the redshift and \texttt{lowT} (hereafter $T_\textrm{low}$) parameters to their minimum values ($10^{-7}$ and 0.0808 keV, respectively). 

Each of these models provides a good description of the broad-band spectrum (Cstat = $266-298$ for 313 dof). The results are recorded in Table \ref{tab: spectral_tableJ1654}, and we display the thermal bremsstrahlung radiation model in Figure \ref{fig: x-ray_spectrumJ1654}. We note that the \texttt{mkcflow} and \texttt{ipolar} models yield consistent measurements for the WD mass. The measurement of the shock temperature $T_\textrm{max}=21\pm3$ keV in the \texttt{mkcflow} is consistent with a WD mass of $M_\textrm{WD}=0.59\pm0.06 M_\odot$ \citep{Mukai2017}, which is the same as determined by \texttt{ipolar} ($M_\textrm{WD}=0.58\pm0.05 M_\odot$). However, if cyclotron cooling is important (for high magnetic field strengths; $B>20$ MG), then these models underestimate the WD mass. We therefore treat these measurements as a lower limit to the true WD mass. We do note, however, that our analysis in Section \ref{sec: sedsection} suggests a low magnetic field strength.

We note that there is a hint of residuals in the $6-7$ keV region of the Fe lines (see Figure \ref{fig: x-ray_spectrumJ1654}), which can be reduced with the addition of a Gaussian emission line ($\texttt{gauss}$) with fixed energy $6.7$ keV and fixed line width $\sigma=0$ keV. Using the bremsstrahlung model, we set an equivalent width upper limit of $<210$ eV. This constraint is within the range of reasonable values for other mCVs \citep{Romanus2015}.

We performed a similar analysis of the \textit{XMM-Newton} data (PN/MOS1/MOS2) obtained in the low-state. The data are sparse with almost no source photons at $>3$ keV, and, therefore, we choose to only apply the power-law and bremsstrahlung models. 
We obtain a best fit for an absorbed power-law (Cstat=816 for 824 dof) with a low Galactic hydrogen column density $N_H=(8\pm5)\times 10^{20}$ cm$^{-2}$ and hard photon index $\Gamma=1.6\pm0.2$. We derive an unabsorbed flux of $(8.8\pm1.3)\times10^{-14}$ erg cm$^{-2}$ s$^{-1}$ in the $0.3-10$ keV energy range, which is a factor of $\sim30\times$ less than the flux observed with XRT.
The low-state spectrum yields a smaller hydrogen column density, consistent with increased absorption during the high-state, possibly due to the accretion column. We performed a test by applying the best-fit high-state model (only allowing the normalization to vary), and found that the increased absorption in the high-state led to a model which under-predicted the low-state spectrum below $1$ keV (consistent with a smaller absorption in the low-state).
We further note that there is no requirement of a soft X-ray excess, generally described by a blackbody component with $T_{bb}<100$ eV \citep{Ramsay2001,Ramsay2004}. 

\begin{table*}
    \caption{Results of our X-ray spectral fitting. The high-state data includes our \textit{Swift} and \textit{NuSTAR} (FPMA and FPMB) spectra, while the low-state includes only the archival \textit{XMM-Newton} (PN, MOS1, and MOS2) data.}
    \label{tab: spectral_tableJ1654}
    \begin{tabular}{lcccccc}
    \hline
    \hline
    &  & \multicolumn{4}{c}{\textbf{Models}} \\
    \cmidrule(lr){3-6}
    \textbf{Parameter} & \textbf{Units} &  \textbf{\texttt{pow}} & \textbf{\texttt{bremss}}  & \textbf{\texttt{mkcflow}} & \textbf{\texttt{ipolar}} \\
    \hline
    \multicolumn{6}{c}{\textbf{High-state}}\\
    \hline
    $N_H$ & $10^{22}$ cm$^{-2}$ & $0.49\pm0.07$ & $0.24\pm0.05$ & $0.35\pm0.05$ & $0.28\pm0.05$ \\
      $\Gamma$  & --& $2.02\pm0.07$ & -- & -- & -- \\
      $N_\texttt{pow}$  & $10^{-4}$ ph cm$^{-2}$ s$^{-1}$ & $8.1\pm1.3$ & -- &  -- & -- \\
     $kT$ & keV & --& $10.1\pm1.2$ & -- & -- \\
     $N_\texttt{bremss}$ & $10^{-4}$ &-- & $5.4\pm0.4$ & --& --\\
     $T_\textrm{high}$ & keV &-- & --& $21\pm3$ & --\\ 
     $A^a$ & -- & --&-- &  $0.8\pm0.3$& -- \\
      $N_\texttt{mkcflow}$ & $10^{-12} M_\odot$ yr$^{-1}$ &-- &-- & $20\pm4 $ & -- \\
    $M_\textrm{WD}$ & $M_\odot$ &-- &-- &-- & $0.58\pm0.06$ \\
    $R_m/R_\textrm{WD}$ & -- &-- &-- &-- & 1000 \\
    $N_\texttt{ipolar}$ & $10^{-28}$ &-- &-- &-- & $5.0\pm1.7$\\
    $C_\textrm{FPMA}$ & -- & 1.0 & 1.0 & 1.0& 1.0 \\
    $C_\textrm{FPMB}$ & -- &$0.94\pm0.07$ &$0.94\pm0.07$ & $0.94\pm0.07$& $0.94\pm0.07$ \\
    $C_\textrm{XRT}$ &-- & $0.63\pm0.07$ & $0.76\pm0.07$ &  $0.72\pm0.08$& $0.75\pm0.08$ \\
      Cstat &-- & 298/313 & 268/313  & 266/312 & 273/313 \\
        \hline
    \multicolumn{6}{c}{\textbf{Low-state}}\\
    \hline
    $N_H$ & $10^{22}$ cm$^{-2}$ & $(8\pm5)\times10^{-2}$ & $(5\pm3)\times10^{-2}$ & -- & -- \\
      $\Gamma$  &-- & $1.6\pm0.2$ & -- & -- & -- \\
      $N_\texttt{pow}$  & $10^{-4}$ ph cm$^{-2}$ s$^{-1}$ & $0.12\pm0.02$ & --&  -- & --\\
          $kT$ & keV &-- & $8^{+9}_{-3}$ & --& --\\
     $N_\texttt{bremss}$ & $10^{-4}$ &-- & $0.15\pm0.02$ & --& --\\
    $C_\textrm{PN}$ & -- & 1.0 & 1.0 & --& --\\
    $C_\textrm{MOS1}$ &-- & $0.84\pm0.18$ & $0.82\pm0.16$ &-- & --\\
    $C_\textrm{MOS2}$ &-- & $1.05\pm0.20$ & $1.04\pm0.19$ &  -- & --\\
     Cstat/dof &-- & 816/824 & 816/824 & -- & --\\
   \hline
   \hline
   \end{tabular}
    \tablecomments{ $^a$ Abundance relative to solar in the \texttt{mkcflow} model.}
\end{table*}

\section{Discussion}
\label{sec: discussionJ1654}

Polars are often luminous soft X-ray sources. The number of known polars expanded greatly thanks to \textit{ROSAT} (e.g., \citealt{Burwitz1997,Burwitz1998,Reinsch1999}). However, their systematic identification effort was limited to high Galactic latitude for practical reasons \citep{Thomas1998,Schwope2002}, among which is the presence of many coronal (i.e., soft) sources dominating the $0.1-2.5$ keV \textit{ROSAT} band at low latitudes. First discovered by \textit{ROSAT}, J1654 was identified through DGPS observations as an unclassified, relatively hard source coincident with a bright \textit{Gaia} star. Based on these qualifiers, we observed the source with a variety of X-ray and optical instruments to aid in its classification. J1654 is yet another example of an X-ray ($0.3-10$ keV) selected polar CV, initially discovered with \textit{ROSAT} in soft X-rays, with a classification confirmed through optical spectroscopy and photometry \citep[e.g.,][]{Thomas1998,Schwope2002,Beuermann2017,Beuermann2020,Beuermann2021}. However, J1654 lies at low Galactic latitude, $b\approx0.01$ deg, demonstrating the importance of X-ray surveys, such as the \textit{Swift} DGPS, with broader energy coverage that can discriminate polars from coronal X-ray sources based on their hardness ratios.

The properties of J1654 outlined in the previous sections are characteristic of polars. The X-ray ($0.3-10$ keV) luminosity varies between $L_X=(6.5\pm0.8)\times10^{31}$ and $(2.3\pm0.4)\times10^{30}$ erg s$^{-1}$ in the high-state and low-state, respectively, assuming a distance of 460 pc. 
These luminosities suggest an accretion rate in the range of $(1-2)\times10^{-11} M_\odot$ yr$^{-1}$. At these distances, the hard X-ray luminosity from \textit{Swift}/BAT is less than $<2\times10^{34}$ erg s$^{-1}$ in the $14-195$ keV energy band. This is consistent with the range of hard X-ray luminosities of known polars \citep{Suleimanov2022}.

In the low-state, the the hydrogen column density $N_{H,\textrm{low}}=(8\pm5)\times 10^{20}$ cm$^{-2}$ yields a Galactic extinction of $A_V=0.36\pm0.23$ mag \citep{Guver2009}. This is consistent with the low value of $A_V\approx 0.13-0.35$ mag \citep{Amores2005} derived from Galactic dust maps, but is at odds with the best fitting SED (see \S \ref{sec: sedsection}). However, X-ray spectra obtained during the high-state yield hydrogen column densities in the range $N_{H,\textrm{high}}=(2-5)\times 10^{21}$ cm$^{-2}$, which leads to a significantly larger $A_V\approx 1-2$ mag \citep{Guver2009}. We therefore cannot rule out that there is significant extinction intrinsic to the source environment (e.g., a circumbinary disk). Further soft X-ray ($0.3-10$ keV) observations, preferably during the high-state, with \textit{Swift}, \textit{XMM-Newton}, or \textit{eROSITA} would be required to better constrain the column density.

Our analysis of the low-state SED implies a secondary star of late-spectral type, consistent with the CV classification. We identify two main possibilities: \textit{i}) a K dwarf with a large extinction $A_V\sim4.9$ mag and \textit{ii}) an M dwarf with a smaller extinction $A_V\sim0.5$ mag. On balance we prefer the latter scenario for multiple reasons. Firstly, the M dwarf scenario has better agreement with the inferred extinction values using the low-state X-ray spectra ($N_{H,\textrm{low}}$), as well as Galactic dust maps. Secondly, a smaller value of the extinction has better agreement with Balmer decrements of order unity in our optical spectra, matching predictions for CVs. Thirdly, interpreting the $UVW1$ detection in the low-state as emission from a WD leads to an estimate of a WD temperature $T=12,000$ K, which is consistent with typical temperatures of accreting WDs \citep{Szkody2010,Pala2017}.

Our classification of the source as a polar agrees with the interpretation of \citet{Takata2022}, who classified the source as a candidate polar, though they did not exclude an IP classification. In this work, we further exclude the possibility that the source is an IP and solidify the polar classification for the following reasons. First, both our analysis and that of \citet{Takata2022} identify only a single period of 2.87 hr. In the polar interpretation, this period is $P_\textrm{orb}=P_\textrm{spin}$. However, if the system is interpreted as an IP then the spin period must be less than the orbital period, and the 2.87 hr period must be one or the other. While 2.87 hr can be the orbital period of an IP, the phase-folded TESS light curve presented by \citet{Takata2022} (their Figure 17) shows a primary maximum lasting half a cycle, with a sharp rise and sharp decline. This is atypical compared to other IPs. Given that much of optical light of an IP is from the partial accretion disk, which is more or less axially symmetric, there is no ready explanation for such an orbital modulation. In fact, optical photometric variability on the orbital period is usually of low amplitude in IPs, and is more gradual in waveform, unless there is an eclipse, to the point that the large majority of orbital period determinations are determined from optical spectroscopy \citep[for examples of the period determinations in IPs see, e.g.,][]{Falomo1987,Hellier1992,Haberl1994,Remillard1994,Allan1999,Norton2002,Zharikov2002,deMartino2006,Scargini2011,Aungwerojwit2012,Halpern2018,Gorgone2021}.

We must also consider the possibility that the 2.87 hr period is the spin period of the WD. This conflicts with known IP systems, where all but 3 of the 71 confirmed IPs\footnote{\url{https://asd.gsfc.nasa.gov/Koji.Mukai/iphome/catalog/alpha.html}} have spin periods shorter than 1 hr. Therefore, this interpretation would make J1654 the longest spin period IP, significantly longer than the
previous record holder (RX J2015.6+3711 at 7196 s, or
just under 2 hrs; \citealt{Halpern2015,CotiZelati2016}). Note that RX J2015.6+3711 has an orbital
period of 12.76 hr and that the spectral features of the secondary
are readily detectable in optical spectra during its normal
(high accretion rate) state \citep{Halpern2018}.
In general, if J1654 is an IP with a 2.87 hr spin period, this would
likely require an orbital period significantly longer than a few hours. 
In contrast, the low-state SED indicates a small mass donor
consistent with the size of the Roche lobe for a $\sim3$ hr orbital
period CV \citep{Knigge2011}. Therefore, we conclude
that the interpretation of J1654 as a polar is the most probable
explanation, and that it matches all available information for the source.

With an orbital period of 2.87 hr, J1654 lies within the CV period gap. This is a common property of polars, which do not exhibit a prominent period gap \citep{Wickramasinghe1994,Webbink2002}. This is thought to be due to reduced magnetic breaking affecting the evolution of polars, see, e.g., \citet{Li1994,Belloni2020}. We discuss the validity of this interpretation for J1654 in more detail below.

We are confident that, if the infrared excess from \textit{Spitzer} is due to cyclotron radiation, the magnetic field has to be rather low ($B\approx 3.5$ MG). This is also supported by the fact that larger magnetic field strengths begin to push the excess towards the optical, where we would expect to observe cyclotron bumps in our SALT spectra (e.g., Figure \ref{fig: spectraJ1654}). 

Furthermore, based on the X-ray luminosity we infer an accretion rate of $\sim$\,$10^{-11} M_\odot$ yr$^{-1}$. While some adjustment is needed to account for the accretion power that goes into cyclotron emission, the correction is likely small due to the fact that we do not detect a luminous soft X-ray component \citep{Ramsay2004}. On the other hand, the inferred rate is based on a single measurement during a high-state, and the long-term average (over high- and low-states) is likely somewhat, but not greatly, lower. This is because Figure \ref{fig: J1654_nustar_lc} suggests that this polar is usually found in a high-state (4 out of 5 epochs since 2007; grouping XRT observations of a similar time into the same epoch).  This should be compared to the number of 21 out of 37 polars found in a high-state by \citet{Ramsay2004other}.

In any case, the accretion rate is comparable to the expected values for systems whose evolution is driven purely by gravitational wave radiation \citep[e.g.,][]{Knigge2011}. This is consistent with the reduced magnetic braking model for the evolution of polars \citep{Li1994,Belloni2020}, in which the magnetic braking mechanism is ineffective for polars due to the WD magnetic field. However, there is a tension between this conclusion and the low magnetic field ($B\approx3.5$ MG) that was obtained by modeling the infrared excess as cyclotron emission: can such a low field nevertheless reduce magnetic braking, and if so, should we not see a similar effect in IPs? 

Further optical and infrared observations, including infrared spectroscopy, and a more secure determination of the magnetic field, is highly desirable. In particular, optical and infrared spectroscopy to identify the phase-dependent cyclotron humps are required before a detailed modeling of the WD mass, accretion rate, and magnetic field can be performed. Infrared spectroscopy may either confirm or exclude cyclotron emission as the nature of the IR excess, and is therefore strongly encouraged.

\section{Conclusions}
\label{sec: conclusionsJ1654}

We have presented the results of the DGPS multi-wavelength follow-up campaign of 1RXS J165424.6-433758 using \textit{Swift}, \textit{NuSTAR}, SAAO, and SALT. The source displays a hard X-ray spectrum in the high-state characterized by thermal bremsstrahlung radiation with temperature $kT=10.1\pm1.2$ keV and luminosity $\sim6.5\times10^{31}$ erg s$^{-1}$. The low-state luminosity, revealed by archival \textit{XMM-Newton} data, is a factor of $\sim30\times$ fainter. 

Moreover, we identify the ultraviolet, optical, and infrared counterpart using archival \textit{Chandra} observations. A SALT spectrum of the source revealed a variety of emission lines of hydrogen and helium, and demonstrated an inverse Balmer decrement. High-speed photometry with the 1m SAAO over multiple nights in 2022 May and June allowed us to derive an orbital period of 2.87 hr. The X-ray, ultraviolet, optical, infrared, and, in particular, emission line, properties allow us to classify J1654 as a polar. This classification is in agreement with the independent analysis of \citet{Takata2022}.

\acknowledgments

The authors thank the anonymous referee for their careful review and constructive feedback. 
B.~O. would like to acknowledge the significant contribution of Nicholas Gorgone to DGPS operations and analyses during his dissertation research. B.~O. acknowledges useful discussions with Jeremy Hare. B.~O. and C.~K. acknowledge support from NASA Grant 80NSSC20K0389. 

This work is based on observations obtained with the Southern African Large Telescope under programme 2021-2-LSP-001.
This work made use of data supplied by the UK \textit{Swift} Science Data Centre at the University of Leicester. This research has made use of the XRT Data Analysis Software (XRTDAS) developed under the responsibility of the ASI Science Data Center (ASDC), Italy. This work made use of data from the NuSTAR mission, a project led by the California Institute of Technology, managed by the Jet Propulsion Laboratory, and funded by the National Aeronautics and Space Administration. 
This research has made use of the NuSTAR Data Analysis Software (NuSTARDAS) jointly developed by the ASI Space Science Data Center (SSDC, Italy) and the California Institute of Technology (Caltech, USA). 
This research has made use of data and/or software provided by the High Energy Astrophysics Science Archive Research Center (HEASARC), which is a service of the Astrophysics Science Division at NASA/GSFC. The scientific results reported in this article are based on observations made by the Chandra X-ray Observatory. 
This research has made use of data obtained from the 4XMM XMM-Newton serendipitous stacked source catalogue 4XMM-DR11s compiled by the institutes of the XMM-Newton Survey Science Centre selected by ESA.

%

\facilities{\textit{Swift}/XRT, \textit{Swift}/UVOT, \textit{NuSTAR}, \textit{XMM-Newton}, SALT, SAAO}


\software{\texttt{HEASoft 6.29c}, \texttt{XRTDAS}, \texttt{NuSTARDAS}, \texttt{SAS}, \texttt{XSPEC} \citep{Arnaud1996}, \texttt{IRAF} \citep{Tody1986}, Astropy \citep{astropy}, \texttt{ARIADNE} \citep{ariadne}, \texttt{dynesty} \citep{dynesty}}




\bibliography{magnetar-bib}{}

\begin{thebibliography}{}
\expandafter\ifx\csname natexlab\endcsname\relax\def\natexlab#1{#1}\fi
\providecommand{\url}[1]{\href{#1}{#1}}
\providecommand{\dodoi}[1]{doi:~\href{http://doi.org/#1}{\nolinkurl{#1}}}
\providecommand{\doeprint}[1]{\href{http://ascl.net/#1}{\nolinkurl{http://ascl.net/#1}}}
\providecommand{\doarXiv}[1]{\href{https://arxiv.org/abs/#1}{\nolinkurl{https://arxiv.org/abs/#1}}}

\bibitem[{{Allan} {et~al.}(1999){Allan}, {Hellier}, \& {Buckley}}]{Allan1999}
{Allan}, A., {Hellier}, C., \& {Buckley}, D. A.~H. 1999, in Astronomical
  Society of the Pacific Conference Series, Vol. 157, Annapolis Workshop on
  Magnetic Cataclysmic Variables, ed. C.~{Hellier} \& K.~{Mukai}, 57

\bibitem[{{Am{\^o}res} \& {L{\'e}pine}(2005)}]{Amores2005}
{Am{\^o}res}, E.~B., \& {L{\'e}pine}, J.~R.~D. 2005, \aj, 130, 659,
  \dodoi{10.1086/430957}

\bibitem[{{Am{\^o}res} {et~al.}(2021){Am{\^o}res}, {Jesus}, {Moitinho},
  {Arsenijevic}, {Levenhagen}, {Marshall}, {Kerber}, {K{\"u}nzel}, \&
  {Moura}}]{Amores2021}
{Am{\^o}res}, E.~B., {Jesus}, R.~M., {Moitinho}, A., {et~al.} 2021, \mnras,
  508, 1788, \dodoi{10.1093/mnras/stab2248}

\bibitem[{{Anderson} {et~al.}(2014){Anderson}, {Gaensler}, {Kaplan}, {Slane},
  {Muno}, {Posselt}, {Hong}, {Murray}, {Steeghs}, {Brogan}, {Drake}, {Farrell},
  {Benjamin}, {Chakrabarty}, {Drew}, {Finley}, {Grindlay}, {Lazio}, {Lee},
  {Mauerhan}, \& {van Kerkwijk}}]{Anderson2014}
{Anderson}, G.~E., {Gaensler}, B.~M., {Kaplan}, D.~L., {et~al.} 2014, \apjs,
  212, 13, \dodoi{10.1088/0067-0049/212/1/13}

\bibitem[{{Arnaud}(1996)}]{Arnaud1996}
{Arnaud}, K.~A. 1996, in Astronomical Society of the Pacific Conference Series,
  Vol. 101, Astronomical Data Analysis Software and Systems V, ed. G.~H.
  {Jacoby} \& J.~{Barnes}, 17

\bibitem[{{Astropy Collaboration} {et~al.}(2013){Astropy Collaboration},
  {Robitaille}, {Tollerud}, {Greenfield}, {Droettboom}, {Bray}, {Aldcroft},
  {Davis}, {Ginsburg}, {Price-Whelan}, {Kerzendorf}, {Conley}, {Crighton},
  {Barbary}, {Muna}, {Ferguson}, {Grollier}, {Parikh}, {Nair}, {Unther},
  {Deil}, {Woillez}, {Conseil}, {Kramer}, {Turner}, {Singer}, {Fox}, {Weaver},
  {Zabalza}, {Edwards}, {Azalee Bostroem}, {Burke}, {Casey}, {Crawford},
  {Dencheva}, {Ely}, {Jenness}, {Labrie}, {Lim}, {Pierfederici}, {Pontzen},
  {Ptak}, {Refsdal}, {Servillat}, \& {Streicher}}]{astropy}
{Astropy Collaboration}, {Robitaille}, T.~P., {Tollerud}, E.~J., {et~al.} 2013,
  \aap, 558, A33, \dodoi{10.1051/0004-6361/201322068}

\bibitem[{{Aungwerojwit} {et~al.}(2012){Aungwerojwit}, {G{\"a}nsicke},
  {Wheatley}, {Pyrzas}, {Staels}, {Krajci}, \&
  {Rodr{\'\i}guez-Gil}}]{Aungwerojwit2012}
{Aungwerojwit}, A., {G{\"a}nsicke}, B.~T., {Wheatley}, P.~J., {et~al.} 2012,
  \apj, 758, 79, \dodoi{10.1088/0004-637X/758/2/79}

\bibitem[{{Bailer-Jones} {et~al.}(2021){Bailer-Jones}, {Rybizki}, {Fouesneau},
  {Demleitner}, \& {Andrae}}]{Bailer-Jones2021}
{Bailer-Jones}, C.~A.~L., {Rybizki}, J., {Fouesneau}, M., {Demleitner}, M., \&
  {Andrae}, R. 2021, VizieR Online Data Catalog, I/352

\bibitem[{{Bailer-Jones} {et~al.}(2018){Bailer-Jones}, {Rybizki}, {Fouesneau},
  {Mantelet}, \& {Andrae}}]{Bailer-Jones2018}
{Bailer-Jones}, C.~A.~L., {Rybizki}, J., {Fouesneau}, M., {Mantelet}, G., \&
  {Andrae}, R. 2018, \aj, 156, 58, \dodoi{10.3847/1538-3881/aacb21}

\bibitem[{{Bailey}(1981)}]{Bailey1981}
{Bailey}, J. 1981, \mnras, 197, 31, \dodoi{10.1093/mnras/197.1.31}

\bibitem[{{Barthelmy} {et~al.}(2005){Barthelmy}, {Barbier}, {Cummings},
  {Fenimore}, {Gehrels}, {Hullinger}, {Krimm}, {Markwardt}, {Palmer},
  {Parsons}, {Sato}, {Suzuki}, {Takahashi}, {Tashiro}, \&
  {Tueller}}]{Barthelmy2005}
{Barthelmy}, S.~D., {Barbier}, L.~M., {Cummings}, J.~R., {et~al.} 2005, \ssr,
  120, 143, \dodoi{10.1007/s11214-005-5096-3}

\bibitem[{{Baumgartner} {et~al.}(2013){Baumgartner}, {Tueller}, {Markwardt},
  {Skinner}, {Barthelmy}, {Mushotzky}, {Evans}, \& {Gehrels}}]{Baumgartner2013}
{Baumgartner}, W.~H., {Tueller}, J., {Markwardt}, C.~B., {et~al.} 2013, \apjs,
  207, 19, \dodoi{10.1088/0067-0049/207/2/19}

\bibitem[{{Belloni} {et~al.}(2020){Belloni}, {Schreiber}, {Pala},
  {G{\"a}nsicke}, {Zorotovic}, \& {Rodrigues}}]{Belloni2020}
{Belloni}, D., {Schreiber}, M.~R., {Pala}, A.~F., {et~al.} 2020, \mnras, 491,
  5717, \dodoi{10.1093/mnras/stz3413}

\bibitem[{{Benjamin} {et~al.}(2003){Benjamin}, {Churchwell}, {Babler}, {Bania},
  {Clemens}, {Cohen}, {Dickey}, {Indebetouw}, {Jackson}, {Kobulnicky},
  {Lazarian}, {Marston}, {Mathis}, {Meade}, {Seager}, {Stolovy}, {Watson},
  {Whitney}, {Wolff}, \& {Wolfire}}]{Benjamin2003}
{Benjamin}, R.~A., {Churchwell}, E., {Babler}, B.~L., {et~al.} 2003, \pasp,
  115, 953, \dodoi{10.1086/376696}

\bibitem[{{Bernardini} {et~al.}(2012){Bernardini}, {de Martino}, {Falanga},
  {Mukai}, {Matt}, {Bonnet-Bidaud}, {Masetti}, \& {Mouchet}}]{Bernardini2012}
{Bernardini}, F., {de Martino}, D., {Falanga}, M., {et~al.} 2012, \aap, 542,
  A22, \dodoi{10.1051/0004-6361/201219233}

\bibitem[{{Beuermann} {et~al.}(2017){Beuermann}, {Burwitz}, {Reinsch},
  {Schwope}, \& {Thomas}}]{Beuermann2017}
{Beuermann}, K., {Burwitz}, V., {Reinsch}, K., {Schwope}, A., \& {Thomas},
  H.~C. 2017, \aap, 603, A47, \dodoi{10.1051/0004-6361/201730800}

\bibitem[{{Beuermann} {et~al.}(2020){Beuermann}, {Burwitz}, {Reinsch},
  {Schwope}, \& {Thomas}}]{Beuermann2020}
---. 2020, \aap, 634, A91, \dodoi{10.1051/0004-6361/201936626}

\bibitem[{{Beuermann} {et~al.}(2021){Beuermann}, {Burwitz}, {Reinsch},
  {Schwope}, \& {Thomas}}]{Beuermann2021}
---. 2021, \aap, 645, A56, \dodoi{10.1051/0004-6361/202038598}

\bibitem[{{Bowman} \& {Holdsworth}(2019)}]{Bowman2019}
{Bowman}, D.~M., \& {Holdsworth}, D.~L. 2019, \aap, 629, A21,
  \dodoi{10.1051/0004-6361/201935640}

\bibitem[{{Breeveld} {et~al.}(2011){Breeveld}, {Landsman}, {Holland}, {Roming},
  {Kuin}, \& {Page}}]{Breeveld2011}
{Breeveld}, A.~A., {Landsman}, W., {Holland}, S.~T., {et~al.} 2011, in American
  Institute of Physics Conference Series, Vol. 1358, Gamma Ray Bursts 2010, ed.
  J.~E. {McEnery}, J.~L. {Racusin}, \& N.~{Gehrels}, 373--376,
  \dodoi{10.1063/1.3621807}

\bibitem[{{Brinkworth} {et~al.}(2007){Brinkworth}, {Hoard}, {Wachter},
  {Howell}, {Ciardi}, {Szkody}, {Harrison}, {van Belle}, \&
  {Esin}}]{Brinkworth2007}
{Brinkworth}, C.~S., {Hoard}, D.~W., {Wachter}, S., {et~al.} 2007, \apj, 659,
  1541, \dodoi{10.1086/512797}

\bibitem[{{Buccheri} {et~al.}(1983){Buccheri}, {Bennett}, {Bignami}, {Bloemen},
  {Boriakoff}, {Caraveo}, {Hermsen}, {Kanbach}, {Manchester}, {Masnou},
  {Mayer-Hasselwander}, {{\"O}zel}, {Paul}, {Sacco}, {Scarsi}, \&
  {Strong}}]{Buccheri1983}
{Buccheri}, R., {Bennett}, K., {Bignami}, G.~F., {et~al.} 1983, \aap, 128, 245

\bibitem[{{Buckley} {et~al.}(2006){Buckley}, {Swart}, \&
  {Meiring}}]{Buckley2006}
{Buckley}, D. A.~H., {Swart}, G.~P., \& {Meiring}, J.~G. 2006, in Society of
  Photo-Optical Instrumentation Engineers (SPIE) Conference Series, Vol. 6267,
  Society of Photo-Optical Instrumentation Engineers (SPIE) Conference Series,
  ed. L.~M. {Stepp}, 62670Z, \dodoi{10.1117/12.673750}

\bibitem[{{Burgh} {et~al.}(2003){Burgh}, {Nordsieck}, {Kobulnicky}, {Williams},
  {O'Donoghue}, {Smith}, \& {Percival}}]{Burgh2003}
{Burgh}, E.~B., {Nordsieck}, K.~H., {Kobulnicky}, H.~A., {et~al.} 2003, in
  Society of Photo-Optical Instrumentation Engineers (SPIE) Conference Series,
  Vol. 4841, Instrument Design and Performance for Optical/Infrared
  Ground-based Telescopes, ed. M.~{Iye} \& A.~F.~M. {Moorwood}, 1463--1471,
  \dodoi{10.1117/12.460312}

\bibitem[{{Burrows} {et~al.}(2005){Burrows}, {Hill}, {Nousek}, {Kennea},
  {Wells}, {Osborne}, {Abbey}, {Beardmore}, {Mukerjee}, {Short}, {Chincarini},
  {Campana}, {Citterio}, {Moretti}, {Pagani}, {Tagliaferri}, {Giommi},
  {Capalbi}, {Tamburelli}, {Angelini}, {Cusumano}, {Br{\"a}uninger}, {Burkert},
  \& {Hartner}}]{Burrows2005}
{Burrows}, D.~N., {Hill}, J.~E., {Nousek}, J.~A., {et~al.} 2005, \ssr, 120,
  165, \dodoi{10.1007/s11214-005-5097-2}

\bibitem[{{Burwitz} {et~al.}(1997){Burwitz}, {Reinsch}, {Beuermann}, \&
  {Thomas}}]{Burwitz1997}
{Burwitz}, V., {Reinsch}, K., {Beuermann}, K., \& {Thomas}, H.~C. 1997, \aap,
  327, 183

\bibitem[{{Burwitz} {et~al.}(1998){Burwitz}, {Reinsch}, {Schwope}, {Hakala},
  {Beuermann}, {Rousseau}, {Thomas}, {G{\"a}nsicke}, {Piirola}, \&
  {Vilhu}}]{Burwitz1998}
{Burwitz}, V., {Reinsch}, K., {Schwope}, A.~D., {et~al.} 1998, \aap, 331, 262

\bibitem[{{Cardelli} {et~al.}(1989){Cardelli}, {Clayton}, \&
  {Mathis}}]{Cardelli1989}
{Cardelli}, J.~A., {Clayton}, G.~C., \& {Mathis}, J.~S. 1989, \apj, 345, 245,
  \dodoi{10.1086/167900}

\bibitem[{{Cash}(1979)}]{Cash1979}
{Cash}, W. 1979, \apj, 228, 939, \dodoi{10.1086/156922}

\bibitem[{{Castelli} \& {Kurucz}(2003)}]{Castelli2004}
{Castelli}, F., \& {Kurucz}, R.~L. 2003, in Modelling of Stellar Atmospheres,
  ed. N.~{Piskunov}, W.~W. {Weiss}, \& D.~F. {Gray}, Vol. 210, A20.
\newblock \doarXiv{astro-ph/0405087}

\bibitem[{{Coppejans} {et~al.}(2013){Coppejans}, {Gulbis}, {Kotze},
  {Coppejans}, {Worters}, {Woudt}, {Whittal}, {Cloete}, \&
  {Fourie}}]{Coppejans2013}
{Coppejans}, R., {Gulbis}, A.~A.~S., {Kotze}, M.~M., {et~al.} 2013, \pasp, 125,
  976, \dodoi{10.1086/672156}

\bibitem[{{Coti Zelati} {et~al.}(2016){Coti Zelati}, {Rea}, {Campana}, {de
  Martino}, {Papitto}, {Safi-Harb}, \& {Torres}}]{CotiZelati2016}
{Coti Zelati}, F., {Rea}, N., {Campana}, S., {et~al.} 2016, \mnras, 456, 1913,
  \dodoi{10.1093/mnras/stv2803}

\bibitem[{{Crawford} {et~al.}(2010){Crawford}, {Still}, {Schellart}, {Balona},
  {Buckley}, {Dugmore}, {Gulbis}, {Kniazev}, {Kotze}, {Loaring}, {Nordsieck},
  {Pickering}, {Potter}, {Romero Colmenero}, {Vaisanen}, {Williams}, \&
  {Zietsman}}]{Crawford2010}
{Crawford}, S.~M., {Still}, M., {Schellart}, P., {et~al.} 2010, in Society of
  Photo-Optical Instrumentation Engineers (SPIE) Conference Series, Vol. 7737,
  Observatory Operations: Strategies, Processes, and Systems III, ed. D.~R.
  {Silva}, A.~B. {Peck}, \& B.~T. {Soifer}, 773725, \dodoi{10.1117/12.857000}

\bibitem[{{Cropper}(1990)}]{Cropper1990}
{Cropper}, M. 1990, \ssr, 54, 195, \dodoi{10.1007/BF00177799}

\bibitem[{{Cropper} {et~al.}(1999){Cropper}, {Wu}, {Ramsay}, \&
  {Kocabiyik}}]{Cropper1999}
{Cropper}, M., {Wu}, K., {Ramsay}, G., \& {Kocabiyik}, A. 1999, \mnras, 306,
  684, \dodoi{10.1046/j.1365-8711.1999.02570.x}

\bibitem[{{de Martino} {et~al.}(2006){de Martino}, {Bonnet-Bidaud}, {Mouchet},
  {G{\"a}nsicke}, {Haberl}, \& {Motch}}]{deMartino2006}
{de Martino}, D., {Bonnet-Bidaud}, J.~M., {Mouchet}, M., {et~al.} 2006, \aap,
  449, 1151, \dodoi{10.1051/0004-6361:20053877}

\bibitem[{{Drew} {et~al.}(2014){Drew}, {Gonzalez-Solares}, {Greimel}, {Irwin},
  {K{\"u}pc{\"u} Yoldas}, {Lewis}, {Barentsen}, {Eisl{\"o}ffel}, {Farnhill},
  {Martin}, {Walsh}, {Walton}, {Mohr-Smith}, {Raddi}, {Sale}, {Wright},
  {Groot}, {Barlow}, {Corradi}, {Drake}, {Fabregat}, {Frew}, {G{\"a}nsicke},
  {Knigge}, {Mampaso}, {Morris}, {Naylor}, {Parker}, {Phillipps}, {Ruhland},
  {Steeghs}, {Unruh}, {Vink}, {Wesson}, \& {Zijlstra}}]{Drew2014}
{Drew}, J.~E., {Gonzalez-Solares}, E., {Greimel}, R., {et~al.} 2014, \mnras,
  440, 2036, \dodoi{10.1093/mnras/stu394}

\bibitem[{{Drew} {et~al.}(2016){Drew}, {Gonzales-Solares}, {Greimel}, {Irwin},
  {Kupcu Yoldas}, {Lewis}, {Barentsen}, {Eisloffel}, {Farnhill}, {Martin},
  {Walsh}, {Walton}, {Mohr-Smith}, {Raddi}, {Sale}, {Wright}, {Groot},
  {Barlow}, {Corradi}, {Drake}, {Fabregat}, {Frew}, {Gansicke}, {Knigge},
  {Mampaso}, {Morris}, {Naylor}, {Parker}, {Phillipps}, {Ruhland}, {Steeghs},
  {Unruh}, {Vink}, {Wesson}, \& {Zijlstra}}]{Drew2016}
{Drew}, J.~E., {Gonzales-Solares}, E., {Greimel}, R., {et~al.} 2016, VizieR
  Online Data Catalog, II/341

\bibitem[{{Evans} {et~al.}(2010){Evans}, {Primini}, {Glotfelty}, {Anderson},
  {Bonaventura}, {Chen}, {Davis}, {Doe}, {Evans}, {Fabbiano}, {Galle}, {Gibbs},
  {Grier}, {Hain}, {Hall}, {Harbo}, {He}, {Houck}, {Karovska}, {Kashyap},
  {Lauer}, {McCollough}, {McDowell}, {Miller}, {Mitschang}, {Morgan},
  {Mossman}, {Nichols}, {Nowak}, {Plummer}, {Refsdal}, {Rots}, {Siemiginowska},
  {Sundheim}, {Tibbetts}, {Van Stone}, {Winkelman}, \& {Zografou}}]{Evans2010}
{Evans}, I.~N., {Primini}, F.~A., {Glotfelty}, K.~J., {et~al.} 2010, \apjs,
  189, 37, \dodoi{10.1088/0067-0049/189/1/37}

\bibitem[{{Evans} {et~al.}(2009){Evans}, {Beardmore}, {Page}, {Osborne},
  {O'Brien}, {Willingale}, {Starling}, {Burrows}, {Godet}, {Vetere}, {Racusin},
  {Goad}, {Wiersema}, {Angelini}, {Capalbi}, {Chincarini}, {Gehrels}, {Kennea},
  {Margutti}, {Morris}, {Mountford}, {Pagani}, {Perri}, {Romano}, \&
  {Tanvir}}]{Evans2009}
{Evans}, P.~A., {Beardmore}, A.~P., {Page}, K.~L., {et~al.} 2009, \mnras, 397,
  1177, \dodoi{10.1111/j.1365-2966.2009.14913.x}

\bibitem[{{Falomo} {et~al.}(1987){Falomo}, {Bonnet-Bidaud}, {Charles},
  {Maraschi}, {Mouchet}, {Mukai}, {Tanzi}, \& {Treves}}]{Falomo1987}
{Falomo}, R., {Bonnet-Bidaud}, J.~M., {Charles}, P.~A., {et~al.} 1987, \apss,
  131, 631, \dodoi{10.1007/BF00668148}

\bibitem[{{Gaia Collaboration}(2020)}]{gaiaEDR3}
{Gaia Collaboration}. 2020, VizieR Online Data Catalog, I/350

\bibitem[{{Gaia Collaboration} {et~al.}(2016){Gaia Collaboration}, {Brown},
  {Vallenari}, {Prusti}, {de Bruijne}, {Mignard}, {Drimmel}, {Babusiaux},
  {Bailer-Jones}, {Bastian}, {Biermann}, {Evans}, {Eyer}, {Jansen}, {Jordi},
  {Katz}, {Klioner}, {Lammers}, {Lindegren}, {Luri}, {O'Mullane}, {Panem},
  {Pourbaix}, {Randich}, {Sartoretti}, {Siddiqui}, {Soubiran}, {Valette}, {van
  Leeuwen}, {Walton}, {Aerts}, {Arenou}, {Cropper}, {H{\o}g}, {Lattanzi},
  {Grebel}, {Holland}, {Huc}, {Passot}, {Perryman}, {Bramante}, {Cacciari},
  {Casta{\~n}eda}, {Chaoul}, {Cheek}, {De Angeli}, {Fabricius}, {Guerra},
  {Hern{\'a}ndez}, {Jean-Antoine-Piccolo}, {Masana}, {Messineo}, {Mowlavi},
  {Nienartowicz}, {Ord{\'o}{\~n}ez-Blanco}, {Panuzzo}, {Portell}, {Richards},
  {Riello}, {Seabroke}, {Tanga}, {Th{\'e}venin}, {Torra}, {Els},
  {Gracia-Abril}, {Comoretto}, {Garcia-Reinaldos}, {Lock}, {Mercier},
  {Altmann}, {Andrae}, {Astraatmadja}, {Bellas-Velidis}, {Benson}, {Berthier},
  {Blomme}, {Busso}, {Carry}, {Cellino}, {Clementini}, {Cowell}, {Creevey},
  {Cuypers}, {Davidson}, {De Ridder}, {de Torres}, {Delchambre}, {Dell'Oro},
  {Ducourant}, {Fr{\'e}mat}, {Garc{\'\i}a-Torres}, {Gosset}, {Halbwachs},
  {Hambly}, {Harrison}, {Hauser}, {Hestroffer}, {Hodgkin}, {Huckle}, {Hutton},
  {Jasniewicz}, {Jordan}, {Kontizas}, {Korn}, {Lanzafame}, {Manteiga},
  {Moitinho}, {Muinonen}, {Osinde}, {Pancino}, {Pauwels}, {Petit},
  {Recio-Blanco}, {Robin}, {Sarro}, {Siopis}, {Smith}, {Smith}, {Sozzetti},
  {Thuillot}, {van Reeven}, {Viala}, {Abbas}, {Abreu Aramburu}, {Accart},
  {Aguado}, {Allan}, {Allasia}, {Altavilla}, {{\'A}lvarez}, {Alves},
  {Anderson}, {Andrei}, {Anglada Varela}, {Antiche}, {Antoja}, {Ant{\'o}n},
  {Arcay}, {Bach}, {Baker}, {Balaguer-N{\'u}{\~n}ez}, {Barache}, {Barata},
  {Barbier}, {Barblan}, {Barrado y Navascu{\'e}s}, {Barros}, {Barstow},
  {Becciani}, {Bellazzini}, {Bello Garc{\'\i}a}, {Belokurov}, {Bendjoya},
  {Berihuete}, {Bianchi}, {Bienaym{\'e}}, {Billebaud}, {Blagorodnova},
  {Blanco-Cuaresma}, {Boch}, {Bombrun}, {Borrachero}, {Bouquillon}, {Bourda},
  {Bouy}, {Bragaglia}, {Breddels}, {Brouillet}, {Br{\"u}semeister},
  {Bucciarelli}, {Burgess}, {Burgon}, {Burlacu}, {Busonero}, {Buzzi}, {Caffau},
  {Cambras}, {Campbell}, {Cancelliere}, {Cantat-Gaudin}, {Carlucci},
  {Carrasco}, {Castellani}, {Charlot}, {Charnas}, {Chiavassa}, {Clotet},
  {Cocozza}, {Collins}, {Costigan}, {Crifo}, {Cross}, {Crosta}, {Crowley},
  {Dafonte}, {Damerdji}, {Dapergolas}, {David}, {David}, {De Cat}, {de Felice},
  {de Laverny}, {De Luise}, {De March}, {de Martino}, {de Souza}, {Debosscher},
  {del Pozo}, {Delbo}, {Delgado}, {Delgado}, {Di Matteo}, {Diakite},
  {Distefano}, {Dolding}, {Dos Anjos}, {Drazinos}, {Duran}, {Dzigan},
  {Edvardsson}, {Enke}, {Evans}, {Eynard Bontemps}, {Fabre}, {Fabrizio},
  {Faigler}, {Falc{\~a}o}, {Farr{\`a}s Casas}, {Federici}, {Fedorets},
  {Fern{\'a}ndez-Hern{\'a}ndez}, {Fernique}, {Fienga}, {Figueras}, {Filippi},
  {Findeisen}, {Fonti}, {Fouesneau}, {Fraile}, {Fraser}, {Fuchs}, {Gai},
  {Galleti}, {Galluccio}, {Garabato}, {Garc{\'\i}a-Sedano}, {Garofalo},
  {Garralda}, {Gavras}, {Gerssen}, {Geyer}, {Gilmore}, {Girona}, {Giuffrida},
  {Gomes}, {Gonz{\'a}lez-Marcos}, {Gonz{\'a}lez-N{\'u}{\~n}ez},
  {Gonz{\'a}lez-Vidal}, {Granvik}, {Guerrier}, {Guillout}, {Guiraud},
  {G{\'u}rpide}, {Guti{\'e}rrez-S{\'a}nchez}, {Guy}, {Haigron},
  {Hatzidimitriou}, {Haywood}, {Heiter}, {Helmi}, {Hobbs}, {Hofmann}, {Holl},
  {Holland}, {Hunt}, {Hypki}, {Icardi}, {Irwin}, {Jevardat de Fombelle},
  {Jofr{\'e}}, {Jonker}, {Jorissen}, {Julbe}, {Karampelas}, {Kochoska},
  {Kohley}, {Kolenberg}, {Kontizas}, {Koposov}, {Kordopatis}, {Koubsky},
  {Krone-Martins}, {Kudryashova}, {Kull}, {Bachchan}, {Lacoste-Seris}, {Lanza},
  {Lavigne}, {Le Poncin-Lafitte}, {Lebreton}, {Lebzelter}, {Leccia}, {Leclerc},
  {Lecoeur-Taibi}, {Lemaitre}, {Lenhardt}, {Leroux}, {Liao}, {Licata},
  {Lindstr{\o}m}, {Lister}, {Livanou}, {Lobel}, {L{\"o}ffler}, {L{\'o}pez},
  {Lorenz}, {MacDonald}, {Magalh{\~a}es Fernandes}, {Managau}, {Mann},
  {Mantelet}, {Marchal}, {Marchant}, {Marconi}, {Marinoni}, {Marrese},
  {Marschalk{\'o}}, {Marshall}, {Mart{\'\i}n-Fleitas}, {Martino}, {Mary},
  {Matijevi{\v{c}}}, {Mazeh}, {McMillan}, {Messina}, {Michalik}, {Millar},
  {Miranda}, {Molina}, {Molinaro}, {Molinaro}, {Moln{\'a}r}, {Moniez},
  {Montegriffo}, {Mor}, {Mora}, {Morbidelli}, {Morel}, {Morgenthaler},
  {Morris}, {Mulone}, {Muraveva}, {Musella}, {Narbonne}, {Nelemans},
  {Nicastro}, {Noval}, {Ord{\'e}novic}, {Ordieres-Mer{\'e}}, {Osborne},
  {Pagani}, {Pagano}, {Pailler}, {Palacin}, {Palaversa}, {Parsons}, {Pecoraro},
  {Pedrosa}, {Pentik{\"a}inen}, {Pichon}, {Piersimoni}, {Pineau}, {Plachy},
  {Plum}, {Poujoulet}, {Pr{\v{s}}a}, {Pulone}, {Ragaini}, {Rago}, {Rambaux},
  {Ramos-Lerate}, {Ranalli}, {Rauw}, {Read}, {Regibo}, {Reyl{\'e}}, {Ribeiro},
  {Rimoldini}, {Ripepi}, {Riva}, {Rixon}, {Roelens}, {Romero-G{\'o}mez},
  {Rowell}, {Royer}, {Ruiz-Dern}, {Sadowski}, {Sagrist{\`a} Sell{\'e}s},
  {Sahlmann}, {Salgado}, {Salguero}, {Sarasso}, {Savietto}, {Schultheis},
  {Sciacca}, {Segol}, {Segovia}, {Segransan}, {Shih}, {Smareglia}, {Smart},
  {Solano}, {Solitro}, {Sordo}, {Soria Nieto}, {Souchay}, {Spagna}, {Spoto},
  {Stampa}, {Steele}, {Steidelm{\"u}ller}, {Stephenson}, {Stoev}, {Suess},
  {S{\"u}veges}, {Surdej}, {Szabados}, {Szegedi-Elek}, {Tapiador}, {Taris},
  {Tauran}, {Taylor}, {Teixeira}, {Terrett}, {Tingley}, {Trager}, {Turon},
  {Ulla}, {Utrilla}, {Valentini}, {van Elteren}, {Van Hemelryck}, {van
  Leeuwen}, {Varadi}, {Vecchiato}, {Veljanoski}, {Via}, {Vicente}, {Vogt},
  {Voss}, {Votruba}, {Voutsinas}, {Walmsley}, {Weiler}, {Weingrill}, {Wevers},
  {Wyrzykowski}, {Yoldas}, {{\v{Z}}erjal}, {Zucker}, {Zurbach}, {Zwitter},
  {Alecu}, {Allen}, {Allende Prieto}, {Amorim}, {Anglada-Escud{\'e}},
  {Arsenijevic}, {Azaz}, {Balm}, {Beck}, {Bernstein}, {Bigot}, {Bijaoui},
  {Blasco}, {Bonfigli}, {Bono}, {Boudreault}, {Bressan}, {Brown}, {Brunet},
  {Bunclark}, {Buonanno}, {Butkevich}, {Carret}, {Carrion}, {Chemin},
  {Ch{\'e}reau}, {Corcione}, {Darmigny}, {de Boer}, {de Teodoro}, {de Zeeuw},
  {Delle Luche}, {Domingues}, {Dubath}, {Fodor}, {Fr{\'e}zouls}, {Fries},
  {Fustes}, {Fyfe}, {Gallardo}, {Gallegos}, {Gardiol}, {Gebran}, {Gomboc},
  {G{\'o}mez}, {Grux}, {Gueguen}, {Heyrovsky}, {Hoar}, {Iannicola}, {Isasi
  Parache}, {Janotto}, {Joliet}, {Jonckheere}, {Keil}, {Kim}, {Klagyivik},
  {Klar}, {Knude}, {Kochukhov}, {Kolka}, {Kos}, {Kutka}, {Lainey}, {LeBouquin},
  {Liu}, {Loreggia}, {Makarov}, {Marseille}, {Martayan}, {Martinez-Rubi},
  {Massart}, {Meynadier}, {Mignot}, {Munari}, {Nguyen}, {Nordlander}, {Ocvirk},
  {O'Flaherty}, {Olias Sanz}, {Ortiz}, {Osorio}, {Oszkiewicz}, {Ouzounis},
  {Palmer}, {Park}, {Pasquato}, {Peltzer}, {Peralta}, {P{\'e}turaud},
  {Pieniluoma}, {Pigozzi}, {Poels}, {Prat}, {Prod'homme}, {Raison}, {Rebordao},
  {Risquez}, {Rocca-Volmerange}, {Rosen}, {Ruiz-Fuertes}, {Russo}, {Sembay},
  {Serraller Vizcaino}, {Short}, {Siebert}, {Silva}, {Sinachopoulos}, {Slezak},
  {Soffel}, {Sosnowska}, {Strai{\v{z}}ys}, {ter Linden}, {Terrell}, {Theil},
  {Tiede}, {Troisi}, {Tsalmantza}, {Tur}, {Vaccari}, {Vachier}, {Valles}, {Van
  Hamme}, {Veltz}, {Virtanen}, {Wallut}, {Wichmann}, {Wilkinson}, {Ziaeepour},
  \& {Zschocke}}]{Gaia2016}
{Gaia Collaboration}, {Brown}, A.~G.~A., {Vallenari}, A., {et~al.} 2016, \aap,
  595, A2, \dodoi{10.1051/0004-6361/201629512}

\bibitem[{{Gaia Collaboration} {et~al.}(2018){Gaia Collaboration}, {Brown},
  {Vallenari}, {Prusti}, {de Bruijne}, {Babusiaux}, {Bailer-Jones}, {Biermann},
  {Evans}, {Eyer}, {Jansen}, {Jordi}, {Klioner}, {Lammers}, {Lindegren},
  {Luri}, {Mignard}, {Panem}, {Pourbaix}, {Randich}, {Sartoretti}, {Siddiqui},
  {Soubiran}, {van Leeuwen}, {Walton}, {Arenou}, {Bastian}, {Cropper},
  {Drimmel}, {Katz}, {Lattanzi}, {Bakker}, {Cacciari}, {Casta{\~n}eda},
  {Chaoul}, {Cheek}, {De Angeli}, {Fabricius}, {Guerra}, {Holl}, {Masana},
  {Messineo}, {Mowlavi}, {Nienartowicz}, {Panuzzo}, {Portell}, {Riello},
  {Seabroke}, {Tanga}, {Th{\'e}venin}, {Gracia-Abril}, {Comoretto},
  {Garcia-Reinaldos}, {Teyssier}, {Altmann}, {Andrae}, {Audard},
  {Bellas-Velidis}, {Benson}, {Berthier}, {Blomme}, {Burgess}, {Busso},
  {Carry}, {Cellino}, {Clementini}, {Clotet}, {Creevey}, {Davidson}, {De
  Ridder}, {Delchambre}, {Dell'Oro}, {Ducourant},
  {Fern{\'a}ndez-Hern{\'a}ndez}, {Fouesneau}, {Fr{\'e}mat}, {Galluccio},
  {Garc{\'\i}a-Torres}, {Gonz{\'a}lez-N{\'u}{\~n}ez}, {Gonz{\'a}lez-Vidal},
  {Gosset}, {Guy}, {Halbwachs}, {Hambly}, {Harrison}, {Hern{\'a}ndez},
  {Hestroffer}, {Hodgkin}, {Hutton}, {Jasniewicz}, {Jean-Antoine-Piccolo},
  {Jordan}, {Korn}, {Krone-Martins}, {Lanzafame}, {Lebzelter}, {L{\"o}ffler},
  {Manteiga}, {Marrese}, {Mart{\'\i}n-Fleitas}, {Moitinho}, {Mora}, {Muinonen},
  {Osinde}, {Pancino}, {Pauwels}, {Petit}, {Recio-Blanco}, {Richards},
  {Rimoldini}, {Robin}, {Sarro}, {Siopis}, {Smith}, {Sozzetti}, {S{\"u}veges},
  {Torra}, {van Reeven}, {Abbas}, {Abreu Aramburu}, {Accart}, {Aerts},
  {Altavilla}, {{\'A}lvarez}, {Alvarez}, {Alves}, {Anderson}, {Andrei},
  {Anglada Varela}, {Antiche}, {Antoja}, {Arcay}, {Astraatmadja}, {Bach},
  {Baker}, {Balaguer-N{\'u}{\~n}ez}, {Balm}, {Barache}, {Barata}, {Barbato},
  {Barblan}, {Barklem}, {Barrado}, {Barros}, {Barstow}, {Bartholom{\'e}
  Mu{\~n}oz}, {Bassilana}, {Becciani}, {Bellazzini}, {Berihuete}, {Bertone},
  {Bianchi}, {Bienaym{\'e}}, {Blanco-Cuaresma}, {Boch}, {Boeche}, {Bombrun},
  {Borrachero}, {Bossini}, {Bouquillon}, {Bourda}, {Bragaglia}, {Bramante},
  {Breddels}, {Bressan}, {Brouillet}, {Br{\"u}semeister}, {Brugaletta},
  {Bucciarelli}, {Burlacu}, {Busonero}, {Butkevich}, {Buzzi}, {Caffau},
  {Cancelliere}, {Cannizzaro}, {Cantat-Gaudin}, {Carballo}, {Carlucci},
  {Carrasco}, {Casamiquela}, {Castellani}, {Castro-Ginard}, {Charlot},
  {Chemin}, {Chiavassa}, {Cocozza}, {Costigan}, {Cowell}, {Crifo}, {Crosta},
  {Crowley}, {Cuypers}, {Dafonte}, {Damerdji}, {Dapergolas}, {David}, {David},
  {de Laverny}, {De Luise}, {De March}, {de Martino}, {de Souza}, {de Torres},
  {Debosscher}, {del Pozo}, {Delbo}, {Delgado}, {Delgado}, {Di Matteo},
  {Diakite}, {Diener}, {Distefano}, {Dolding}, {Drazinos}, {Dur{\'a}n},
  {Edvardsson}, {Enke}, {Eriksson}, {Esquej}, {Eynard Bontemps}, {Fabre},
  {Fabrizio}, {Faigler}, {Falc{\~a}o}, {Farr{\`a}s Casas}, {Federici},
  {Fedorets}, {Fernique}, {Figueras}, {Filippi}, {Findeisen}, {Fonti},
  {Fraile}, {Fraser}, {Fr{\'e}zouls}, {Gai}, {Galleti}, {Garabato},
  {Garc{\'\i}a-Sedano}, {Garofalo}, {Garralda}, {Gavel}, {Gavras}, {Gerssen},
  {Geyer}, {Giacobbe}, {Gilmore}, {Girona}, {Giuffrida}, {Glass}, {Gomes},
  {Granvik}, {Gueguen}, {Guerrier}, {Guiraud}, {Guti{\'e}rrez-S{\'a}nchez},
  {Haigron}, {Hatzidimitriou}, {Hauser}, {Haywood}, {Heiter}, {Helmi}, {Heu},
  {Hilger}, {Hobbs}, {Hofmann}, {Holland}, {Huckle}, {Hypki}, {Icardi},
  {Jan{\ss}en}, {Jevardat de Fombelle}, {Jonker}, {Juh{\'a}sz}, {Julbe},
  {Karampelas}, {Kewley}, {Klar}, {Kochoska}, {Kohley}, {Kolenberg},
  {Kontizas}, {Kontizas}, {Koposov}, {Kordopatis}, {Kostrzewa-Rutkowska},
  {Koubsky}, {Lambert}, {Lanza}, {Lasne}, {Lavigne}, {Le Fustec}, {Le
  Poncin-Lafitte}, {Lebreton}, {Leccia}, {Leclerc}, {Lecoeur-Taibi},
  {Lenhardt}, {Leroux}, {Liao}, {Licata}, {Lindstr{\o}m}, {Lister}, {Livanou},
  {Lobel}, {L{\'o}pez}, {Managau}, {Mann}, {Mantelet}, {Marchal}, {Marchant},
  {Marconi}, {Marinoni}, {Marschalk{\'o}}, {Marshall}, {Martino}, {Marton},
  {Mary}, {Massari}, {Matijevi{\v{c}}}, {Mazeh}, {McMillan}, {Messina},
  {Michalik}, {Millar}, {Molina}, {Molinaro}, {Moln{\'a}r}, {Montegriffo},
  {Mor}, {Morbidelli}, {Morel}, {Morris}, {Mulone}, {Muraveva}, {Musella},
  {Nelemans}, {Nicastro}, {Noval}, {O'Mullane}, {Ord{\'e}novic},
  {Ord{\'o}{\~n}ez-Blanco}, {Osborne}, {Pagani}, {Pagano}, {Pailler},
  {Palacin}, {Palaversa}, {Panahi}, {Pawlak}, {Piersimoni}, {Pineau}, {Plachy},
  {Plum}, {Poggio}, {Poujoulet}, {Pr{\v{s}}a}, {Pulone}, {Racero}, {Ragaini},
  {Rambaux}, {Ramos-Lerate}, {Regibo}, {Reyl{\'e}}, {Riclet}, {Ripepi}, {Riva},
  {Rivard}, {Rixon}, {Roegiers}, {Roelens}, {Romero-G{\'o}mez}, {Rowell},
  {Royer}, {Ruiz-Dern}, {Sadowski}, {Sagrist{\`a} Sell{\'e}s}, {Sahlmann},
  {Salgado}, {Salguero}, {Sanna}, {Santana-Ros}, {Sarasso}, {Savietto},
  {Schultheis}, {Sciacca}, {Segol}, {Segovia}, {S{\'e}gransan}, {Shih},
  {Siltala}, {Silva}, {Smart}, {Smith}, {Solano}, {Solitro}, {Sordo}, {Soria
  Nieto}, {Souchay}, {Spagna}, {Spoto}, {Stampa}, {Steele},
  {Steidelm{\"u}ller}, {Stephenson}, {Stoev}, {Suess}, {Surdej}, {Szabados},
  {Szegedi-Elek}, {Tapiador}, {Taris}, {Tauran}, {Taylor}, {Teixeira},
  {Terrett}, {Teyssandier}, {Thuillot}, {Titarenko}, {Torra Clotet}, {Turon},
  {Ulla}, {Utrilla}, {Uzzi}, {Vaillant}, {Valentini}, {Valette}, {van Elteren},
  {Van Hemelryck}, {van Leeuwen}, {Vaschetto}, {Vecchiato}, {Veljanoski},
  {Viala}, {Vicente}, {Vogt}, {von Essen}, {Voss}, {Votruba}, {Voutsinas},
  {Walmsley}, {Weiler}, {Wertz}, {Wevers}, {Wyrzykowski}, {Yoldas},
  {{\v{Z}}erjal}, {Ziaeepour}, {Zorec}, {Zschocke}, {Zucker}, {Zurbach}, \&
  {Zwitter}}]{Gaia2018}
---. 2018, \aap, 616, A1, \dodoi{10.1051/0004-6361/201833051}

\bibitem[{{Gorgone} {et~al.}(2021){Gorgone}, {Woudt}, {Buckley}, {Mukai},
  {Kouveliotou}, {Gogus}, {Bellm}, {Linford}, {van der Horst}, {Baring},
  {Hartmann}, {Barrett}, {Cenko}, {Graham}, {Granot}, {Harrison}, {Kennea},
  {O'Connor}, {Potter}, {Stern}, \& {Wijers}}]{Gorgone2021}
{Gorgone}, N.~M., {Woudt}, P.~A., {Buckley}, D., {et~al.} 2021, arXiv e-prints,
  arXiv:2103.14800.
\newblock \doarXiv{2103.14800}

\bibitem[{{G{\"u}ver} \& {{\"O}zel}(2009)}]{Guver2009}
{G{\"u}ver}, T., \& {{\"O}zel}, F. 2009, \mnras, 400, 2050,
  \dodoi{10.1111/j.1365-2966.2009.15598.x}

\bibitem[{{Haberl} {et~al.}(1994){Haberl}, {Throstensen}, {Motch},
  {Schwarzenberg-Czerny}, {Pakull}, {Shambrook}, \& {Pietsch}}]{Haberl1994}
{Haberl}, F., {Throstensen}, J.~R., {Motch}, C., {et~al.} 1994, \aap, 291, 171

\bibitem[{{Halpern} \& {Thorstensen}(2015)}]{Halpern2015}
{Halpern}, J.~P., \& {Thorstensen}, J.~R. 2015, \aj, 150, 170,
  \dodoi{10.1088/0004-6256/150/6/170}

\bibitem[{{Halpern} {et~al.}(2018){Halpern}, {Thorstensen}, {Cho}, {Collver},
  {Motsoaledi}, {Breytenbach}, {Buckley}, \& {Woudt}}]{Halpern2018}
{Halpern}, J.~P., {Thorstensen}, J.~R., {Cho}, P., {et~al.} 2018, \aj, 155,
  247, \dodoi{10.3847/1538-3881/aabfd0}

\bibitem[{{Harrison} \& {Campbell}(2015)}]{Harrison2015}
{Harrison}, T.~E., \& {Campbell}, R.~K. 2015, \apjs, 219, 32,
  \dodoi{10.1088/0067-0049/219/2/32}

\bibitem[{{Hellier} \& {Sproats}(1992)}]{Hellier1992}
{Hellier}, C., \& {Sproats}, L.~N. 1992, Information Bulletin on Variable
  Stars, 3724, 1

\bibitem[{{Hessman} {et~al.}(2000){Hessman}, {G{\"a}nsicke}, \&
  {Mattei}}]{Hessman2000}
{Hessman}, F.~V., {G{\"a}nsicke}, B.~T., \& {Mattei}, J.~A. 2000, \aap, 361,
  952

\bibitem[{Husser {et~al.}(2013)Husser, von Berg, Dreizler, Homeier, Reiners,
  Barman, \& Hauschildt}]{Husser2013}
Husser, T.-O., von Berg, S.~W., Dreizler, S., {et~al.} 2013, Astronomy {\&}
  Astrophysics, 553, A6, \dodoi{10.1051/0004-6361/201219058}

\bibitem[{{Knigge}(2006)}]{Knigge2006}
{Knigge}, C. 2006, \mnras, 373, 484, \dodoi{10.1111/j.1365-2966.2006.11096.x}

\bibitem[{{Knigge} {et~al.}(2011){Knigge}, {Baraffe}, \&
  {Patterson}}]{Knigge2011}
{Knigge}, C., {Baraffe}, I., \& {Patterson}, J. 2011, \apjs, 194, 28,
  \dodoi{10.1088/0067-0049/194/2/28}

\bibitem[{{Kobulnicky} {et~al.}(2003){Kobulnicky}, {Nordsieck}, {Burgh},
  {Smith}, {Percival}, {Williams}, \& {O'Donoghue}}]{Kobulnicky2003}
{Kobulnicky}, H.~A., {Nordsieck}, K.~H., {Burgh}, E.~B., {et~al.} 2003, in
  Society of Photo-Optical Instrumentation Engineers (SPIE) Conference Series,
  Vol. 4841, Instrument Design and Performance for Optical/Infrared
  Ground-based Telescopes, ed. M.~{Iye} \& A.~F.~M. {Moorwood}, 1634--1644,
  \dodoi{10.1117/12.460315}

\bibitem[{{Krimm} {et~al.}(2013){Krimm}, {Holland}, {Corbet}, {Pearlman},
  {Romano}, {Kennea}, {Bloom}, {Barthelmy}, {Baumgartner}, {Cummings},
  {Gehrels}, {Lien}, {Markwardt}, {Palmer}, {Sakamoto}, {Stamatikos}, \&
  {Ukwatta}}]{Krimm2013}
{Krimm}, H.~A., {Holland}, S.~T., {Corbet}, R.~H.~D., {et~al.} 2013, \apjs,
  209, 14, \dodoi{10.1088/0067-0049/209/1/14}

\bibitem[{{Kurucz}(1993)}]{Kurucz1993}
{Kurucz}, R.~L. 1993, VizieR Online Data Catalog, VI/39

\bibitem[{{Li} {et~al.}(1994){Li}, {Wu}, \& {Wickramasinghe}}]{Li1994}
{Li}, J.~K., {Wu}, K.~W., \& {Wickramasinghe}, D.~T. 1994, \mnras, 268, 61,
  \dodoi{10.1093/mnras/268.1.61}

\bibitem[{{Lien} {et~al.}(2016){Lien}, {Sakamoto}, {Barthelmy}, {Baumgartner},
  {Cannizzo}, {Chen}, {Collins}, {Cummings}, {Gehrels}, {Krimm}, {Markwardt},
  {Palmer}, {Stamatikos}, {Troja}, \& {Ukwatta}}]{Lien2016}
{Lien}, A., {Sakamoto}, T., {Barthelmy}, S.~D., {et~al.} 2016, \apj, 829, 7,
  \dodoi{10.3847/0004-637X/829/1/7}

\bibitem[{{Mason} {et~al.}(2001){Mason}, {Breeveld}, {Much}, {Carter},
  {Cordova}, {Cropper}, {Fordham}, {Huckle}, {Ho}, {Kawakami}, {Kennea},
  {Kennedy}, {Mittaz}, {Pandel}, {Priedhorsky}, {Sasseen}, {Shirey}, {Smith},
  \& {Vreux}}]{Mason2001}
{Mason}, K.~O., {Breeveld}, A., {Much}, R., {et~al.} 2001, \aap, 365, L36,
  \dodoi{10.1051/0004-6361:20000044}

\bibitem[{{Meggitt} \& {Wickramasinghe}(1982)}]{Meggitt1982}
{Meggitt}, S.~M.~A., \& {Wickramasinghe}, D.~T. 1982, \mnras, 198, 71,
  \dodoi{10.1093/mnras/198.1.71}

\bibitem[{{Minniti} {et~al.}(2017){Minniti}, {Lucas}, \& {VVV
  Team}}]{Minniti2017}
{Minniti}, D., {Lucas}, P., \& {VVV Team}. 2017, VizieR Online Data Catalog,
  II/348

\bibitem[{{Mukai}(2017)}]{Mukai2017}
{Mukai}, K. 2017, \pasp, 129, 062001, \dodoi{10.1088/1538-3873/aa6736}

\bibitem[{{Mushotzky} \& {Szymkowiak}(1988)}]{Mushotzky1988}
{Mushotzky}, R.~F., \& {Szymkowiak}, A.~E. 1988, in NATO Advanced Study
  Institute (ASI) Series C, Vol. 229, Cooling Flows in Clusters and Galaxies,
  ed. A.~C. {Fabian}, 53, \dodoi{10.1007/978-94-009-2953-1_6}

\bibitem[{{Newville} {et~al.}(2016){Newville}, {Stensitzki}, {Allen}, {Rawlik},
  {Ingargiola}, \& {Nelson}}]{lmfit}
{Newville}, M., {Stensitzki}, T., {Allen}, D.~B., {et~al.} 2016, {Lmfit:
  Non-Linear Least-Square Minimization and Curve-Fitting for Python},
  Astrophysics Source Code Library, record ascl:1606.014.
\newblock \doeprint{1606.014}

\bibitem[{{Norton} {et~al.}(2002){Norton}, {Quaintrell}, {Katajainen}, {Lehto},
  {Mukai}, \& {Negueruela}}]{Norton2002}
{Norton}, A.~J., {Quaintrell}, H., {Katajainen}, S., {et~al.} 2002, \aap, 384,
  195, \dodoi{10.1051/0004-6361:20011820}

\bibitem[{{O'Connor} {et~al.}(2023{\natexlab{a}}){O'Connor}, {Kouveliotou},
  {Evans}, {Gorgone}, {van Kooten}, {Gagnon}, {Yang}, {Baring}, {Bellm},
  {Beniamini}, {Brink}, {Buckley}, {Cenko}, {Egbo}, {Gogus}, {Granot},
  {Hailey}, {Hare}, {Harrison}, {Hartmann}, {van der Horst}, {Huppenkothen},
  {Kaper}, {Kargaltsev}, {Kennea}, {Mukai}, {Slane}, {Stern}, {Troja},
  {Wadiasingh}, {Wijers}, {Woudt}, \& {Younes}}]{DGPS}
{O'Connor}, B., {Kouveliotou}, C., {Evans}, P.~A., {et~al.} 2023{\natexlab{a}},
  arXiv e-prints, arXiv:2306.14354.
\newblock \doarXiv{2306.14354}

\bibitem[{{O'Connor} {et~al.}(2023{\natexlab{b}}){O'Connor}, {Kouveliotou},
  {Evans}, {Gorgone}, {van Kooten}, {Gagnon}, {Yang}, {Baring}, {Bellm},
  {Beniamini}, {Brink}, {Buckley}, {Cenko}, {Egbo}, {Gogus}, {Granot},
  {Hailey}, {Hare}, {Harrison}, {Hartmann}, {van der Horst}, {Huppenkothen},
  {Kaper}, {Kargaltsev}, {Kennea}, {Mukai}, {Slane}, {Stern}, {Troja},
  {Wadiasingh}, {Wijers}, {Woudt}, \& {Younes}}]{J1708}
---. 2023{\natexlab{b}}, arXiv e-prints, arXiv:2306.14354,
  \dodoi{10.48550/arXiv.2306.14354}

\bibitem[{{Oh} {et~al.}(2018){Oh}, {Koss}, {Markwardt}, {Schawinski},
  {Baumgartner}, {Barthelmy}, {Cenko}, {Gehrels}, {Mushotzky}, {Petulante},
  {Ricci}, {Lien}, \& {Trakhtenbrot}}]{Oh2018}
{Oh}, K., {Koss}, M., {Markwardt}, C.~B., {et~al.} 2018, \apjs, 235, 4,
  \dodoi{10.3847/1538-4365/aaa7fd}

\bibitem[{{Oliveira} {et~al.}(2017){Oliveira}, {Rodrigues}, {Cieslinski},
  {Jablonski}, {Silva}, {Almeida}, {Rodr{\'\i}guez-Ardila}, \&
  {Palhares}}]{Oliveira2017}
{Oliveira}, A.~S., {Rodrigues}, C.~V., {Cieslinski}, D., {et~al.} 2017, \aj,
  153, 144, \dodoi{10.3847/1538-3881/aa610d}

\bibitem[{{Oliveira} {et~al.}(2020){Oliveira}, {Rodrigues}, {Martins},
  {Palhares}, {Silva}, {Lima}, \& {Jablonski}}]{Oliveira2020}
{Oliveira}, A.~S., {Rodrigues}, C.~V., {Martins}, M., {et~al.} 2020, \aj, 159,
  114, \dodoi{10.3847/1538-3881/ab6ded}

\bibitem[{{Pala} {et~al.}(2017){Pala}, {G{\"a}nsicke}, {Townsley}, {Boyd},
  {Cook}, {De Martino}, {Godon}, {Haislip}, {Henden}, {Hubeny}, {Ivarsen},
  {Kafka}, {Knigge}, {LaCluyze}, {Long}, {Marsh}, {Monard}, {Moore}, {Myers},
  {Nelson}, {Nogami}, {Oksanen}, {Pickard}, {Poyner}, {Reichart}, {Rodriguez
  Perez}, {Schreiber}, {Shears}, {Sion}, {Stubbings}, {Szkody}, \&
  {Zorotovic}}]{Pala2017}
{Pala}, A.~F., {G{\"a}nsicke}, B.~T., {Townsley}, D., {et~al.} 2017, \mnras,
  466, 2855, \dodoi{10.1093/mnras/stw3293}

\bibitem[{{Pala} {et~al.}(2020){Pala}, {G{\"a}nsicke}, {Breedt}, {Knigge},
  {Hermes}, {Gentile Fusillo}, {Hollands}, {Naylor}, {Pelisoli}, {Schreiber},
  {Toonen}, {Aungwerojwit}, {Cukanovaite}, {Dennihy}, {Manser}, {Pretorius},
  {Scaringi}, \& {Toloza}}]{Pala2020}
{Pala}, A.~F., {G{\"a}nsicke}, B.~T., {Breedt}, E., {et~al.} 2020, \mnras, 494,
  3799, \dodoi{10.1093/mnras/staa764}

\bibitem[{{Parsons} {et~al.}(2018){Parsons}, {G{\"a}nsicke}, {Marsh}, {Ashley},
  {Breedt}, {Burleigh}, {Copperwheat}, {Dhillon}, {Green}, {Hermes}, {Irawati},
  {Kerry}, {Littlefair}, {Rebassa-Mansergas}, {Sahman}, {Schreiber}, \&
  {Zorotovic}}]{Parsons2018}
{Parsons}, S.~G., {G{\"a}nsicke}, B.~T., {Marsh}, T.~R., {et~al.} 2018, \mnras,
  481, 1083, \dodoi{10.1093/mnras/sty2345}

\bibitem[{{Patterson}(1994)}]{Patterson1994}
{Patterson}, J. 1994, \pasp, 106, 209, \dodoi{10.1086/133375}

\bibitem[{{Pottasch}(1984)}]{Pottasch1984}
{Pottasch}, S.~R. 1984, {Planetary nebulae. A study of late stages of stellar
  evolution}, Vol. 107 (Astrophysics and Space Science Library),
  \dodoi{10.1007/978-94-009-7233-9}

\bibitem[{{Potter} {et~al.}(2002){Potter}, {Ramsay}, {Wu}, \&
  {Cropper}}]{Potter2002}
{Potter}, S., {Ramsay}, G., {Wu}, K., \& {Cropper}, M. 2002, in Astronomical
  Society of the Pacific Conference Series, Vol. 261, The Physics of
  Cataclysmic Variables and Related Objects, ed. B.~T. {G{\"a}nsicke},
  K.~{Beuermann}, \& K.~{Reinsch}, 165

\bibitem[{{Ramsay} \& {Cropper}(2004)}]{Ramsay2004}
{Ramsay}, G., \& {Cropper}, M. 2004, \mnras, 347, 497,
  \dodoi{10.1111/j.1365-2966.2004.07220.x}

\bibitem[{{Ramsay} {et~al.}(2001){Ramsay}, {Cropper}, {C{\'o}rdova}, {Mason},
  {Much}, {Pandel}, \& {Shirey}}]{Ramsay2001}
{Ramsay}, G., {Cropper}, M., {C{\'o}rdova}, F., {et~al.} 2001, \mnras, 326,
  L27, \dodoi{10.1046/j.1365-8711.2001.04798.x}

\bibitem[{{Ramsay} {et~al.}(2004){Ramsay}, {Cropper}, {Wu}, {Mason},
  {C{\'o}rdova}, \& {Priedhorsky}}]{Ramsay2004other}
{Ramsay}, G., {Cropper}, M., {Wu}, K., {et~al.} 2004, \mnras, 350, 1373,
  \dodoi{10.1111/j.1365-2966.2004.07732.x}

\bibitem[{{Reinsch} {et~al.}(1999){Reinsch}, {Burwitz}, {Beuermann}, \&
  {Thomas}}]{Reinsch1999}
{Reinsch}, K., {Burwitz}, V., {Beuermann}, K., \& {Thomas}, H.-C. 1999, in
  Astronomical Society of the Pacific Conference Series, Vol. 157, Annapolis
  Workshop on Magnetic Cataclysmic Variables, ed. C.~{Hellier} \& K.~{Mukai},
  187

\bibitem[{{Remillard} {et~al.}(1994){Remillard}, {Bradt}, {Brissenden},
  {Buckley}, {Schwartz}, {Silber}, {Stroozas}, \& {Tuohy}}]{Remillard1994}
{Remillard}, R.~A., {Bradt}, H.~V., {Brissenden}, R.~J.~V., {et~al.} 1994,
  \apj, 428, 785, \dodoi{10.1086/174287}

\bibitem[{{Reynolds} {et~al.}(2013){Reynolds}, {Miller}, {Maitra}, {Gultekin},
  {Gehrels}, {Kennea}, {Siegel}, {Gelbord}, \& {Kuin}}]{Reynolds2013}
{Reynolds}, M.~T., {Miller}, J.~M., {Maitra}, D., {et~al.} 2013, The
  Astronomer's Telegram, 5200, 1

\bibitem[{{Romanus} {et~al.}(2015){Romanus}, {Saitou}, \&
  {Ebisawa}}]{Romanus2015}
{Romanus}, E., {Saitou}, K., \& {Ebisawa}, K. 2015, arXiv e-prints,
  arXiv:1511.09424, \dodoi{10.48550/arXiv.1511.09424}

\bibitem[{{Roming} {et~al.}(2005){Roming}, {Kennedy}, {Mason}, {Nousek}, {Ahr},
  {Bingham}, {Broos}, {Carter}, {Hancock}, {Huckle}, {Hunsberger}, {Kawakami},
  {Killough}, {Koch}, {McLelland}, {Smith}, {Smith}, {Soto}, {Boyd},
  {Breeveld}, {Holland}, {Ivanushkina}, {Pryzby}, {Still}, \&
  {Stock}}]{Roming2005}
{Roming}, P. W.~A., {Kennedy}, T.~E., {Mason}, K.~O., {et~al.} 2005, \ssr, 120,
  95, \dodoi{10.1007/s11214-005-5095-4}

\bibitem[{{Sakamoto} {et~al.}(2011){Sakamoto}, {Barthelmy}, {Baumgartner},
  {Cummings}, {Fenimore}, {Gehrels}, {Krimm}, {Markwardt}, {Palmer}, {Parsons},
  {Sato}, {Stamatikos}, {Tueller}, {Ukwatta}, \& {Zhang}}]{Sakamoto2011}
{Sakamoto}, T., {Barthelmy}, S.~D., {Baumgartner}, W.~H., {et~al.} 2011, \apjs,
  195, 2, \dodoi{10.1088/0067-0049/195/1/2}

\bibitem[{{Scargle}(1982)}]{Scargle1982}
{Scargle}, J.~D. 1982, \apj, 263, 835, \dodoi{10.1086/160554}

\bibitem[{{Scaringi} {et~al.}(2011){Scaringi}, {Connolly}, {Patterson},
  {Thorstensen}, {Uthas}, {Knigge}, {Vican}, {Monard}, {Rea}, {Krajci},
  {Lowther}, {Myers}, {Bolt}, {Dieball}, \& {Groot}}]{Scargini2011}
{Scaringi}, S., {Connolly}, S., {Patterson}, J., {et~al.} 2011, \aap, 530, A6,
  \dodoi{10.1051/0004-6361/201116585}

\bibitem[{{Schwope} {et~al.}(2002){Schwope}, {Brunner}, {Buckley}, {Greiner},
  {Heyden}, {Neizvestny}, {Potter}, \& {Schwarz}}]{Schwope2002}
{Schwope}, A.~D., {Brunner}, H., {Buckley}, D., {et~al.} 2002, \aap, 396, 895,
  \dodoi{10.1051/0004-6361:20021386}

\bibitem[{{Schwope} {et~al.}(1997){Schwope}, {Buckley}, {O'Donoghue},
  {Hasinger}, {Truemper}, \& {Voges}}]{Schwope1997}
{Schwope}, A.~D., {Buckley}, D.~A.~H., {O'Donoghue}, D., {et~al.} 1997, \aap,
  326, 195.
\newblock \doarXiv{astro-ph/9705106}

\bibitem[{{Shaw} {et~al.}(2020){Shaw}, {Heinke}, {Mukai}, {Tomsick},
  {Doroshenko}, {Suleimanov}, {Buisson}, {Gandhi}, {Grefenstette}, {Hare},
  {Jiang}, {Ludlam}, {Rana}, \& {Sivakoff}}]{Shaw2020}
{Shaw}, A.~W., {Heinke}, C.~O., {Mukai}, K., {et~al.} 2020, \mnras, 498, 3457,
  \dodoi{10.1093/mnras/staa2592}

\bibitem[{{Silber}(1992)}]{Silber1992}
{Silber}, A.~D. 1992, PhD thesis, Massachusetts Institute of Technology

\bibitem[{{Smith} {et~al.}(2018){Smith}, {Lucas}, {Kurtev}, {Smart}, {Minniti},
  {Borissova}, {Jones}, {Zhang}, {Marocco}, {Contreras Pe{\~n}a}, {Gromadzki},
  {Kuhn}, {Drew}, {Pinfield}, \& {Bedin}}]{Smith2018}
{Smith}, L.~C., {Lucas}, P.~W., {Kurtev}, R., {et~al.} 2018, \mnras, 474, 1826,
  \dodoi{10.1093/mnras/stx2789}

\bibitem[{{Smith} {et~al.}(2020){Smith}, {Lucas}, {Kurtev}, {Smart}, {Minniti},
  {Borissova}, {Jones}, {Zhang}, {Marocco}, {Contreras Pena}, {Gromadzki},
  {Kuhn}, {Drew}, {Pinfield}, \& {Bedin}}]{Smith2020}
---. 2020, VizieR Online Data Catalog, II/364

\bibitem[{{Smith} {et~al.}(2006){Smith}, {Nordsieck}, {Burgh}, {Percival},
  {Williams}, {O'Donohue}, {O'Connor}, \& {Schier}}]{Smith2006}
{Smith}, M.~P., {Nordsieck}, K.~H., {Burgh}, E.~B., {et~al.} 2006, in Society
  of Photo-Optical Instrumentation Engineers (SPIE) Conference Series, Vol.
  6269, Society of Photo-Optical Instrumentation Engineers (SPIE) Conference
  Series, ed. I.~S. {McLean} \& M.~{Iye}, 62692A, \dodoi{10.1117/12.672415}

\bibitem[{{Speagle}(2020)}]{dynesty}
{Speagle}, J.~S. 2020, \mnras, 493, 3132, \dodoi{10.1093/mnras/staa278}

\bibitem[{{Stockman} {et~al.}(1977){Stockman}, {Schmidt}, {Angel}, {Liebert},
  {Tapia}, \& {Beaver}}]{Stockman1977}
{Stockman}, H.~S., {Schmidt}, G.~D., {Angel}, J.~R.~P., {et~al.} 1977, \apj,
  217, 815, \dodoi{10.1086/155629}

\bibitem[{{Suleimanov} {et~al.}(2016){Suleimanov}, {Doroshenko}, {Ducci},
  {Zhukov}, \& {Werner}}]{Suleimanov2016}
{Suleimanov}, V., {Doroshenko}, V., {Ducci}, L., {Zhukov}, G.~V., \& {Werner},
  K. 2016, \aap, 591, A35, \dodoi{10.1051/0004-6361/201628301}

\bibitem[{{Suleimanov} {et~al.}(2019){Suleimanov}, {Doroshenko}, \&
  {Werner}}]{Suleimanov2019}
{Suleimanov}, V.~F., {Doroshenko}, V., \& {Werner}, K. 2019, \mnras, 482, 3622,
  \dodoi{10.1093/mnras/sty2952}

\bibitem[{{Suleimanov} {et~al.}(2022){Suleimanov}, {Doroshenko}, \&
  {Werner}}]{Suleimanov2022}
---. 2022, \mnras, 511, 4937, \dodoi{10.1093/mnras/stac417}

\bibitem[{{Szkody} {et~al.}(2010){Szkody}, {Mukadam}, {G{\"a}nsicke}, {Henden},
  {Templeton}, {Holtzman}, {Montgomery}, {Howell}, {Nitta}, {Sion}, {Schwartz},
  \& {Dillon}}]{Szkody2010}
{Szkody}, P., {Mukadam}, A., {G{\"a}nsicke}, B.~T., {et~al.} 2010, \apj, 710,
  64, \dodoi{10.1088/0004-637X/710/1/64}

\bibitem[{{Szkody} {et~al.}(2020){Szkody}, {Dicenzo}, {Ho}, {Hillenbrand}, {van
  Roestel}, {Ridder}, {DeJesus Lima}, {Graham}, {Bellm}, {Burdge}, {Kupfer},
  {Prince}, {Masci}, {Mr{\'o}z}, {Golkhou}, {Coughlin}, {Cunningham}, {Dekany},
  {Graham}, {Hale}, {Kaplan}, {Kasliwal}, {Miller}, {Neill}, {Patterson},
  {Riddle}, {Smith}, \& {Soumagnac}}]{Szkody2020}
{Szkody}, P., {Dicenzo}, B., {Ho}, A. Y.~Q., {et~al.} 2020, \aj, 159, 198,
  \dodoi{10.3847/1538-3881/ab7cce}

\bibitem[{{Takata} {et~al.}(2022){Takata}, {Wang}, {Kong}, {Mao}, {Hou}, {Hu},
  {Lin}, {Li}, \& {Hui}}]{Takata2022}
{Takata}, J., {Wang}, X.~F., {Kong}, A.~K.~H., {et~al.} 2022, \apj, 936, 134,
  \dodoi{10.3847/1538-4357/ac8100}

\bibitem[{{Thomas} {et~al.}(1998){Thomas}, {Beuermann}, {Reinsch}, {Schwope},
  {Truemper}, \& {Voges}}]{Thomas1998}
{Thomas}, H.~C., {Beuermann}, K., {Reinsch}, K., {et~al.} 1998, \aap, 335, 467

\bibitem[{{Tody}(1986)}]{Tody1986}
{Tody}, D. 1986, in Society of Photo-Optical Instrumentation Engineers (SPIE)
  Conference Series, Vol. 627, Instrumentation in astronomy VI, ed. D.~L.
  {Crawford}, 733, \dodoi{10.1117/12.968154}

\bibitem[{{Tonry} {et~al.}(2021){Tonry}, {Denneau}, {Flewelling}, {Heinze},
  {Onken}, {Smartt}, {Stalder}, {Weiland}, \& {Wolf}}]{Tonry2021}
{Tonry}, J.~L., {Denneau}, L., {Flewelling}, H., {et~al.} 2021, VizieR Online
  Data Catalog, J/ApJ/867/105

\bibitem[{{Traulsen} {et~al.}(2020){Traulsen}, {Schwope}, {Lamer}, {Ballet},
  {Carrera}, {Ceballos}, {Coriat}, {Freyberg}, {Koliopanos}, {Kurpas},
  {Michel}, {Motch}, {Page}, {Watson}, \& {Webb}}]{Traulsen2020}
{Traulsen}, I., {Schwope}, A.~D., {Lamer}, G., {et~al.} 2020, \aap, 641, A137,
  \dodoi{10.1051/0004-6361/202037706}

\bibitem[{{Vanderplas}(2015)}]{Gatspy}
{Vanderplas}, J. 2015, {Gatspy: General Tools For Astronomical Time Series In
  Python}, v0.1.1, Zenodo,  Zenodo, \dodoi{10.5281/zenodo.14833}

\bibitem[{{Verner} {et~al.}(1996){Verner}, {Ferland}, {Korista}, \&
  {Yakovlev}}]{Verner1996}
{Verner}, D.~A., {Ferland}, G.~J., {Korista}, K.~T., \& {Yakovlev}, D.~G. 1996,
  \apj, 465, 487, \dodoi{10.1086/177435}

\bibitem[{{Vines} \& {Jenkins}(2022)}]{ariadne}
{Vines}, J.~I., \& {Jenkins}, J.~S. 2022, \mnras, \dodoi{10.1093/mnras/stac956}

\bibitem[{{Warner}(1987)}]{Warner1987}
{Warner}, B. 1987, \mnras, 227, 23, \dodoi{10.1093/mnras/227.1.23}

\bibitem[{{Warner}(1995)}]{Warner1995}
---. 1995, {Cataclysmic variable stars}, Vol.~28

\bibitem[{{Warner}(1999)}]{Warner1999}
{Warner}, B. 1999, in Astronomical Society of the Pacific Conference Series,
  Vol. 157, Annapolis Workshop on Magnetic Cataclysmic Variables, ed.
  C.~{Hellier} \& K.~{Mukai}, 63

\bibitem[{{Warner}(2003)}]{Warner2003}
---. 2003, {Cataclysmic Variable Stars} (Cambridge University Press),
  \dodoi{10.1017/CBO9780511586491}

\bibitem[{{Webbink} \& {Wickramasinghe}(2002)}]{Webbink2002}
{Webbink}, R.~F., \& {Wickramasinghe}, D.~T. 2002, \mnras, 335, 1,
  \dodoi{10.1046/j.1365-8711.2002.05495.x}

\bibitem[{{Wickramasinghe} \& {Meggitt}(1985)}]{Wickramasinghe1985}
{Wickramasinghe}, D.~T., \& {Meggitt}, S.~M.~A. 1985, \mnras, 214, 605,
  \dodoi{10.1093/mnras/214.4.605}

\bibitem[{{Wickramasinghe} \& {Wu}(1994)}]{Wickramasinghe1994}
{Wickramasinghe}, D.~T., \& {Wu}, K. 1994, \mnras, 266, L1,
  \dodoi{10.1093/mnras/266.1.L1}

\bibitem[{{Williams}(1980)}]{Williams1980}
{Williams}, R.~E. 1980, \apj, 235, 939, \dodoi{10.1086/157698}

\bibitem[{{Williams}(1989)}]{Williams1989}
---. 1989, \aj, 97, 1752, \dodoi{10.1086/115115}

\bibitem[{{Williams} \& {Ferguson}(1982)}]{Williams1982}
{Williams}, R.~E., \& {Ferguson}, D.~H. 1982, \apj, 257, 672,
  \dodoi{10.1086/160022}

\bibitem[{{Wilms} {et~al.}(2000){Wilms}, {Allen}, \& {McCray}}]{Wilms2000}
{Wilms}, J., {Allen}, A., \& {McCray}, R. 2000, \apj, 542, 914,
  \dodoi{10.1086/317016}

\bibitem[{{Wu} \& {Kiss}(2008)}]{Wu2008}
{Wu}, K., \& {Kiss}, L.~L. 2008, \aap, 481, 433,
  \dodoi{10.1051/0004-6361:20078556}

\bibitem[{{Zharikov} {et~al.}(2002){Zharikov}, {Tovmassian}, \&
  {Echevarr{\'\i}a}}]{Zharikov2002}
{Zharikov}, S.~V., {Tovmassian}, G.~H., \& {Echevarr{\'\i}a}, J. 2002, \aap,
  390, L23, \dodoi{10.1051/0004-6361:20020913}

\end{thebibliography}
\bibliographystyle{aasjournal}



\appendix
\setcounter{table}{0}
\renewcommand{\thetable}{A\arabic{table}}

\section{Log of Observations}

Here we present tables reporting the log of X-ray (Table \ref{tab: observationsJ1654}), ultraviolet (Table \ref{tab: observationsUVOT}), and optical (SALT; Table \ref{tab: observationsSALT}) observations in this work.

\begin{table*}[ht]
\centering
\caption{
Log of X-ray observations used in this work. 
}
\label{tab: observationsJ1654}
\begin{tabular}{lccccc}
\hline\hline
\textbf{Start Time (UT)} & \textbf{Telescope} & \textbf{Instrument} &  \textbf{Exposure (s)} & \textbf{ObsID}  \\
\hline\hline
 2007-10-23 15:22:50& \textit{Chandra}  & HRC-I  & 1590 & 8231 \\
 2011-01-21 23:47:59 & \textit{Swift} & XRT  & 496 & 00043184001 \\
2012-08-20 21:58:28 & \textit{XMM-Newton} & MOS1/MOS2/PN   & 14916  & 0695000301 \\
2020-06-18 06:40:36 & \textit{Swift} & XRT  & 3941 & 03110780002$^a$ \\
 2020-07-14 12:10:35& \textit{Swift} & XRT  & 1003 & 03110780003$^a$ \\
 2020-07-31 13:30:36& \textit{Swift} & XRT  & 1727  & 00089169001$^a$ \\
 2020-08-02 03:54:59 & \textit{NuSTAR}  & FPMA/B   &  25655 & 90601324002$^b$ \\
 2022-04-04 03:17:00   & \textit{Swift} & XRT  & 855 & 00043184003$^b$ \\
 2022-04-07 02:58:00  & \textit{Swift} & XRT  & 950 & 00043184004$^b$ \\
 2022-04-12 11:56:00  & \textit{Swift} & XRT  & 990 & 00043184005$^b$ \\
 2022-04-17 16:04:00 & \textit{Swift} & XRT  & 760 & 00043184006$^b$ \\
\hline\hline
\end{tabular}
\tablecomments{ $^a$DGPS observation, $^b$DGPS ToO.} 
\end{table*}

\begin{table*}[ht]
\centering
\caption{
Log of Ultraviolet observations of J1654. }
\label{tab: observationsUVOT}
\begin{tabular}{lccccccc}
\hline\hline
\textbf{Start Time (UT)} &  \textbf{Telescope} & \textbf{Instrument} &   \textbf{ObsID} & \textbf{Filter} & \textbf{Exposure (s)} & \textbf{AB mag}  \\
\hline\hline
 2011-01-21 23:49:44 & \textit{Swift} & UVOT  &  00043184001 & \textit{uvm2} & 494 & $19.77\pm0.11$ \\
  2012-08-20 22:07:32 & \textit{XMM-Newton} & OM & 0695000301 & $V$ & 1500 & $>19.4$ \\
  2012-08-20 22:37:39  & \textit{XMM-Newton} & OM & 0695000301 & $B$ & 1900 & $>20.9$ \\
 2012-08-20 23:14:26 & \textit{XMM-Newton} & OM & 0695000301 & $UVW1$ &5000 & $21.5\pm0.2$ \\
 2012-08-21 00:42:53    & \textit{XMM-Newton} & OM & 0695000301 & $UVW2$ & 5000 & $>21.2$ \\
2020-06-18 06:46:54  & \textit{Swift} & UVOT  &  03110780002& \textit{uvw1} & 301 & $19.27\pm0.12$ \\
2020-06-18 13:16:15  & \textit{Swift} & UVOT  &  03110780002& \textit{uvw1} & 820 & $19.24\pm0.07$  \\
2020-06-18 14:27:56  & \textit{Swift} & UVOT  &  03110780002& \textit{uvw1} & 899 & $19.34\pm0.07$ \\
2020-06-18 17:59:39  & \textit{Swift} & UVOT  &  03110780002& \textit{uvw1} & 676 & $19.45\pm0.08$ \\
2020-06-18 21:01:41  & \textit{Swift} & UVOT  &  03110780002& \textit{uvw1} & 674 & $19.22\pm0.07$ \\
2020-06-18 22:49:21  & \textit{Swift} & UVOT  &  03110780002& \textit{uvw1} & 521 & $19.42\pm0.09$ \\
 2020-07-14 12:15:02 & \textit{Swift} & UVOT  & 03110780003&\textit{uvw1} & 475 & $19.33\pm0.10$ \\
  2020-07-14 18:37:13 & \textit{Swift} & UVOT  & 03110780003& \textit{uvw1} & 524 & $19.36\pm0.10$ \\
 2020-07-31 13:35:09 & \textit{Swift} & UVOT  &  00089169001& \textit{uvw2}  & 1726 & $19.91\pm0.07$  \\
 2022-04-04 03:19:46  & \textit{Swift} & UVOT  & 00043184003& \textit{uvw2} & 850 & $19.67\pm0.08$ \\
 2022-04-07 02:59:10 & \textit{Swift} & UVOT  & 00043184004 & \textit{u} & 946& $18.72\pm0.10$\\
 2022-04-12 11:59:50  & \textit{Swift} & UVOT  &  00043184005  &\textit{uvw2} & 186& $20.21\pm0.24$ \\
 2022-04-12 13:37:41  & \textit{Swift} & UVOT  &  00043184005 &\textit{uvw2}& 434 & $19.80\pm0.12$ \\
 2022-04-12 15:16:01  & \textit{Swift} & UVOT  &  00043184005 &\textit{uvw2}& 354 & $19.94\pm0.15$ \\
 2022-04-17 16:05:13 & \textit{Swift} & UVOT  &  00043184006 &\textit{uvm2} & 523 & $19.84\pm0.12$\\
 2022-04-17 22:38:07 & \textit{Swift} & UVOT  &  00043184006 & \textit{uvm2}& 229 & $19.90\pm0.19$ \\
 \hline\hline
\end{tabular}
\end{table*}

\begin{table*}[ht]
\centering
\caption{
Log of SALT and SAAO observations of J1654. The modes refer to longslit (LS) spectroscopy, frame-transfer (FT) spectroscopy, or high-speed photometry (HSP) with an exposure time of either 5 or 10 s. 
}
\label{tab: observationsSALT}
\begin{tabular}{lcccccc}
\hline\hline
\textbf{Start Time (UT)} &  \textbf{Telescope} & \textbf{Mode} &   \textbf{Grating} & \textbf{Filter} & \textbf{Duration (hr)}   \\
\hline\hline
\multicolumn{6}{c}{\textbf{Optical Spectroscopy}}\\
\hline
2022-04-26 22:12:39 &  SALT & LS  & PG0300  &  -- & 0.42  \\
	2022-05-02 21:24:20 &  SALT & FT & PG0900  & -- & 0.83  \\
	2022-05-10 03:05:48 &  SALT & FT & PG0900  &-- &  0.83 \\
 	2022-05-27 02:10:09 &  SALT & FT & PG0900  &-- & 0.70  \\
\hline
\multicolumn{6}{c}{\textbf{High-speed Optical Photometry}}\\
\hline 
2022-05-04 21:33:00 &  SAAO & HSP - 5 s & --  & \textit{clear} & 6.9  \\
 2022-05-05 21:07:00&  SAAO & HSP - 5 s & --  & \textit{clear} & 6.3   \\
2022-05-08 20:54:00 &  SAAO & HSP - 5 s & --  & $i'$ & 6.9  \\
2022-05-09 20:45:00 &  SAAO & HSP - 10 s &  -- &$g'$ &  7.1 \\
 2022-06-01 19:42:00 &  SAAO & HSP - 10 s &  -- & $i'$ & 8.4  \\
2022-06-02 19:39:00   &  SAAO & HSP -  5 s &  -- & $r'$ & 8.3  \\
 2022-06-07 22:18:00   &  SAAO & HSP - 10 s &  -- & $g'$ & 4.1  \\
  2022-06-08 20:07:00   &  SAAO & HSP -  10 s &  -- &$i'$ & 3.1  \\
 \hline\hline
\end{tabular}
\end{table*}

\section{Secondary Star Modeling}

Here we show the corner plot derived using \texttt{ARIADNE} \citep{ariadne}. The best fit model is displayed in Figure \ref{fig: sedJ1654}.

\begin{figure}
\centering
\includegraphics[width=1\columnwidth]{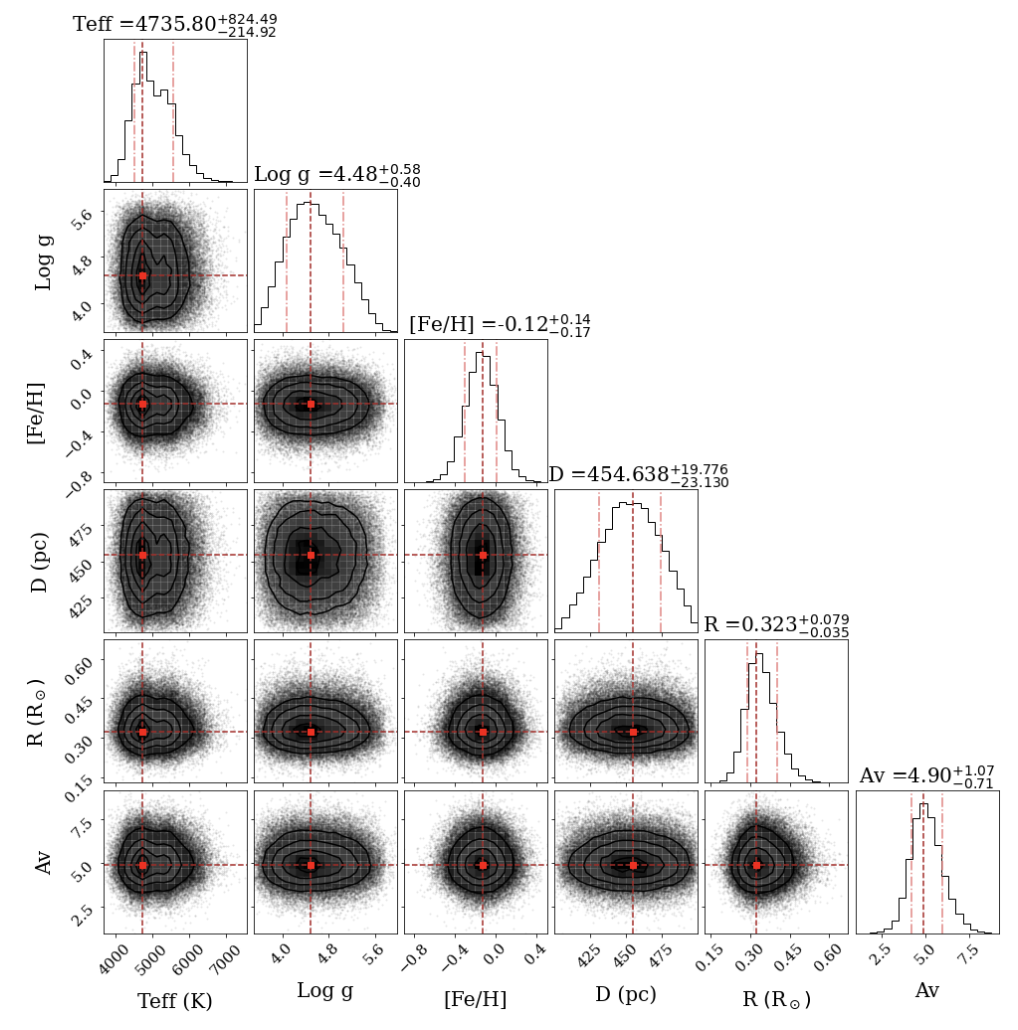}
\caption{Corner plot of secondary star parameters derived from modeling the low-state SED with \texttt{ARIADNE}.
 }
\label{fig: ariadne}
\end{figure}

\end{document}